\documentclass[fleqn,usenatbib]{mnras}

\usepackage{newtxtext,newtxmath}

\usepackage[T1]{fontenc}

\DeclareRobustCommand{\VAN}[3]{#2}
\let\VANthebibliography\thebibliography
\def\thebibliography{\DeclareRobustCommand{\VAN}[3]{##3}\VANthebibliography}

\usepackage{graphicx}	
\usepackage{amsmath}	
\usepackage[compact]{titlesec}
\usepackage{subcaption,ulem}
\usepackage{hyperref}
\hypersetup{
    colorlinks=true,
    linkcolor=blue,
    filecolor=magenta,      
    urlcolor=cyan,
}

\title[Multiphase gas in the cloud-crushing problem with cooling]{Growth and structure of multiphase gas in the cloud-crushing problem with cooling}
\author[Kanjilal, Dutta, Sharma 2020]{
Vijit Kanjilal,$^{1}$\thanks{E-mail: vijitk@iisc.ac.in}
Alankar Dutta,$^{1}$\thanks{E-mail: alankardutta@iisc.ac.in}
Prateek Sharma$^{1}$\thanks{E-mail: prateek@iisc.ac.in}
\thanks{Visualizations related to this work are hosted in a playlist of our \href{https://www.youtube.com/watch?v=TO68aRBFO5I&list=PLH4Yu6hxblq3scjLE7U7EvTQtebIpn_TA}{IISc Computational Astrophysics} YouTube channel.}
\\
$^{1}$Department of Physics, Indian Institute of Science, Bangalore, Karnataka, India
}

\date{Accepted XXX. Received YYY; in original form ZZZ}

\pubyear{2020}

\begin{document}
\label{firstpage}
\pagerange{\pageref{firstpage}--\pageref{lastpage}}
\maketitle

\defcitealias{gronke2018}{GO18}
\defcitealias{gronke2020}{GO20}
\defcitealias{Li2020}{Li+20}

\begin{abstract}
We revisit the problem of the growth of dense/cold gas in the cloud-crushing setup with radiative cooling. The relative motion between the dense cloud and the diffuse medium produces a turbulent boundary layer of mixed gas with a short cooling time. This mixed gas may explain the ubiquity of the range of absorption/emission lines observed in various sources such as the circumgalactic medium and galactic/stellar/AGN outflows. Recently Gronke \& Oh showed that the efficient radiative cooling of the mixed gas can lead to the continuous growth of the dense cloud. They presented a threshold cloud size for the growth of dense gas which was contradicted by the more recent works of Li et al. \& Sparre et al. These thresholds are qualitatively different as the former is based on the cooling time of the mixed gas whereas the latter is based on the cooling time of the hot gas. Our simulations agree with the threshold based on the cooling time of the mixed gas. We argue that the radiative cloud-crushing simulations should be run long enough to allow for the late-time growth of the dense gas due to cooling of the mixed gas but not so long that the background gas cools catastrophically. Moreover, the simulation domain should be large enough that the mixed gas is not lost through the boundaries. While the mixing layer is roughly isobaric, the emissivity of the gas at different temperatures is fundamentally different from an isobaric single-phase steady cooling flow.
\end{abstract}



\section{Introduction}
\label{sec:introduction}
Multiphase plasma, with a broad range of temperatures and densities but with rough pressure balance, are very common in various astrophysical systems, ranging from clusters of galaxies (\citealt{molendi2001,sharma2018}), the circumgalactic medium (CGM; \citealt{tumlinson2017}), galactic/AGN outflows (\citealt{veilleux2020,perna2017}), the interstellar medium (\citealt{SVG2002,Audit2005}), to the lower solar corona (\citealt{antolin2020}). While these systems vary greatly in their scales, several of the physical processes have to be common across these diverse environments (e.g., see \citealt{sharma2013}). This motivates the study of simple, idealized physical setups in which we try to identify general physical principles and apply them to make sense of observations (qualitatively, to begin with) and more sophisticated numerical simulations that include several physical elements. One such idealized setup is the hydrodynamic cloud-crushing problem in which a dilute wind blows over a {\it pre-existing} dense cloud, and the interaction between the two is studied (\citealt{klein1994}). Cold gas moving relative to a hot ambient medium is expected generically and therefore it is important to study this interaction in detail. It is also well known that in the absence of radiative cooling the cloud is mixed in the diffuse wind much before it can be pushed by it (e.g., \citealt{zhang2017}); this is true even with magnetic fields and thermal conduction (e.g., \citealt{BB2016,cottle2020,bruggen2016,armillotta2017}). 

Recently, \citet{gronke2018} (hereafter \citetalias{gronke2018}) pointed out that in the presence of radiative cooling, sufficiently large clouds can entrain material from the hot wind and grow rather than get mixed in the diffuse phase (see also \citealt{mellema2002,armillotta2016,gritton2017}). They also derived an analytic criterion for the growth of cold cloud. Even more recently, \citet{Li2020} (hereafter \citetalias{Li2020}) and \citet{Sparre2020} have presented a criterion for  cloud growth that differs qualitatively and quantitatively from \citetalias{gronke2018}. The present work aims to independently study this important problem.

While our results are broadly applicable, we choose parameters relevant for the CGM. Recent observations have revealed the multiphase structure of the CGM (\citealt{tumlinson2017}). The cold phase primarily consists of neutral and low ionization potential ions like H I, Na I, Ca II, and dust; the cool phase  harbours ions like C II, C III, Si II, Si III, N II, and N III;  the warm phase is traced by high ionization potential ions like C IV, N V, O VI, and Ne VII; and  the hot gas phase by highly ionized species like O VII and O VIII. The absorption (and less commonly emission) features of these different species serve as tracers of the multiphase CGM. The CGM is polluted by metals due to stellar winds and supernovae, which can coalesce at a sufficiently high star formation rate density (\citealt{heckman2001,yadav2017}), forming a large-scale galactic wind. Star formation distributed through the disk can often result in entrained cold gas moving at high velocity (\citealt{cooper2008,vijayan2018,schneider2020}), as inferred from spectral line observations (for a review, see \citealt{veilluex2005,rupke2018}). Multiphase gas is seen not only in galactic outflows, but also in the extended CGM as probed by quasar absorption (\citealt{tumlinson2011,werk2014}). Most likely, this multiphase gas with a sub-escape velocity and spreading isotropically across the halo over $\sim 100$ kpc is not associated with outflows but arises mostly due to condensation induced near wakes of satellite galaxies and intergalactic filaments (\citealt{choudhury2019,nelson2020,mandelker2020,Fielding2020b}).  Cold gas is also seen in cool core clusters within 10s of kpc of the cluster center (e.g., \citealt{olivares2019}) and extended symmetrically out to the virial radius in $z\gtrsim 3$ halos hosting quasars (e.g., \citealt{borisova2016}), and within 10s of kpc in $z\gtrsim 3$  star-forming galaxies (e.g., \citealt{wisotzki2018}).

The observations of cold gas moving up to several 100 km s$^{-1}$ raise a major question on the existing theoretical models. It is expected that hydrodynamic instabilities, such as the Kelvin-Helmholtz and Rayleigh-Taylor instabilities, should mix the cold cloud into the diffuse hot phase and ultimately destroy it. The destruction timescale of an adiabatic, initially static cloud of size $R_{\rm cl}$ (in pressure equilibrium with its background; Table~\ref{tab:Variable_defn} lists the various relevant parameters for the cloud-crushing problem) exposed to an impinging hot wind of velocity $v_{\rm wind}$ scales as the cloud-crushing time $t_{\rm cc}\sim {\sqrt{\chi} R_{\rm cl}}/{v_{\rm wind}}$ (where $\chi$ is the density contrast between the cloud and the wind; \citealt{klein1994}). The timescale of the wind to accelerate the cloud via momentum transfer to speed $\sim v_{\rm wind}$ is given by the drag time $t_{\rm drag}\sim \chi {R_{\rm cl}}/{v_{\rm wind}}$. Clearly, the cloud-crushing time $t_{\rm cc}$ is shorter by a factor of $\sqrt{\chi }$ than the drag time $t_{\rm drag}$, implying that the cloud must be destroyed before it is blown away by the wind. This problem -- the survival of cool gas clouds moving through a hot ambient medium, the cloud-crushing problem, -- has received significant attention in recent years, particularly in the context of the CGM (\citealt{BB2015,schneider2017a,BB2018,gronke2018,gronke2020,Sparre2018,Liang2019,Li2020,Sparre2020}).

\begin{table}
	\centering
	\caption{Parameters for the cloud-crushing problem}
	\label{tab:Variable_defn}
	\begin{tabular}{lc} 
		\hline
		Symbol & Meaning\\
		\hline
		$R_{\rm cl}$ & Cloud radius (diameter $L_{\rm cl}=2R_{\rm cl}$)\\
		$d_{\rm cell}$   & Cell/grid size in simulation (resolution) \\
		$n_{\rm cl}$ & Particle number density in the cloud\\
		$n_{\rm hot}$ & Particle number density in the wind\\
		$T_{\rm cl}$ & Temperature of the dense cloud (equals cooling floor)\\
		$T_{\rm hot}$ & Temperature of the hot wind\\
		$T_{\rm mix}$ & Temperature of the mixed phase, taken to be $\sqrt{T_{\rm hot}T_{\rm cl}}$\\
		$\chi$ & Wind-cloud contrast $(n_{\rm cl}/n_{\rm hot}=T_{\rm hot}/T_{\rm cl})$ \\
		$v_{\rm wind}$ & Relative speed of the hot wind with respect to the cloud\\
		$t_{\rm cc}$ & Classical cloud-crushing time ($\sqrt{\chi}R_{\rm cl}/v_{\rm wind}$)\\
		$t_{\rm drag}$ & Drag time ($\chi R_{\rm cl}/v_{\rm wind}$)\\
		$\mathcal{M}$ & $\equiv v_{\rm wind}/c_s$; $c_s=\sqrt{1.67P/\rho}$ is hot wind sound speed\\
		$t_{\rm cool,hot}$ & Cooling time of the hot phase\\
		$t_{\rm cool,mix}$ & Cooling time of the mixed phase\\
		\hline
	\end{tabular}
\end{table}

Recently, \citetalias{gronke2018}, \citet{gronke2020} (hereafter \citetalias{gronke2020}) revisited the problem of entrainment of hot wind by a cold cloud and derived a criterion for cloud growth due to radiative cooling. They concluded that whenever the ratio of the cooling time of the {\it mixed} gas and the cloud-crushing time ($t_{\rm cool,mix}/t_{\rm cc})< 1$, the mixed warm gas in the boundary layer\footnote{Throughout this paper we use terms such as 'turbulent boundary layer' and 'cloud wake' interchangeably to refer to the mixed gas at the interface of the cloud and the wind.} cools and produces new comoving cold gas, and thus, the cloud grows and acquires momentum. More recently, however, \citetalias{Li2020} derived a criterion for cold mass growth due to radiative cooling and it is somewhat different from the predictions of \citetalias{gronke2018}. According to them, cloud growth occurs only if $t_{\rm cool,hot}/t_{\rm life,pred}< 1$, where $t_{\rm cool,hot}$ is the cooling time of the {\it hot} (not {\it mixed} as in \citetalias{gronke2018}) phase and $t_{\rm life,pred}$ is a predicted cloud lifetime obtained by fitting their simulation results to a simple function of the various parameters of the cloud-crushing problem. They estimate the predicted cloud lifetime as  
\begin{equation}
\label{eq:tlife_pred}
t_{\rm life,pred} \approx 10t_{\rm cc}\overline{f},
\end{equation} where
\begin{equation}
\label{eq:fbar}
\overline{f} = (0.9\pm 0.1)L_{1}^{0.3}n_{0.01}^{0.3}T_{6}^{0.0} v_{100}^{0.6},
\end{equation}
and $L_{1}\equiv L_{\rm cl}/(1$ pc) ($L_{\rm cl}$ is the initial cloud diameter), $n_{0.01}\equiv n_{\rm hot}/(0.01\: {\rm~cm}^{-3}$), $T_{6}\equiv T_{\rm hot}/(10^6$ K) and $v_{100}\equiv v_{\rm wind}/(100\: {\rm~km}{\rm~s}^{-1}$). A very similar criterion was used by \citet{Sparre2020}.

In this paper, we therefore seek to resolve this discrepancy through a set of  three-dimensional idealized hydrodynamical simulations with radiative cooling, similar to that of \citetalias{gronke2018}, \citetalias{gronke2020} and \citetalias{Li2020}, with the simulation parameters specifically chosen to address this issue. We also provide some novel insights into the physics of the radiative cloud-crushing problem. The paper is organized as follows. In section~\ref{sec:analytic} we present the relevant analytical calculations to highlight the discrepancy in terms of the initial cloud size. In section~\ref{sec:num} we describe the simulation details and the range of parameters surveyed in our simulations. In section~\ref{sec:results} we present our results and discuss their implications. We also present the different diagnostics 
to better illustrate the underlying physics. In section~\ref{sec:caveats} we present caveats and future directions. Finally in section~\ref{sec:summary}, we summarize our results.

\section{Analytic Arguments}

\begin{figure*}
\begin{subfigure}{\textwidth}
  \includegraphics[width=\linewidth]{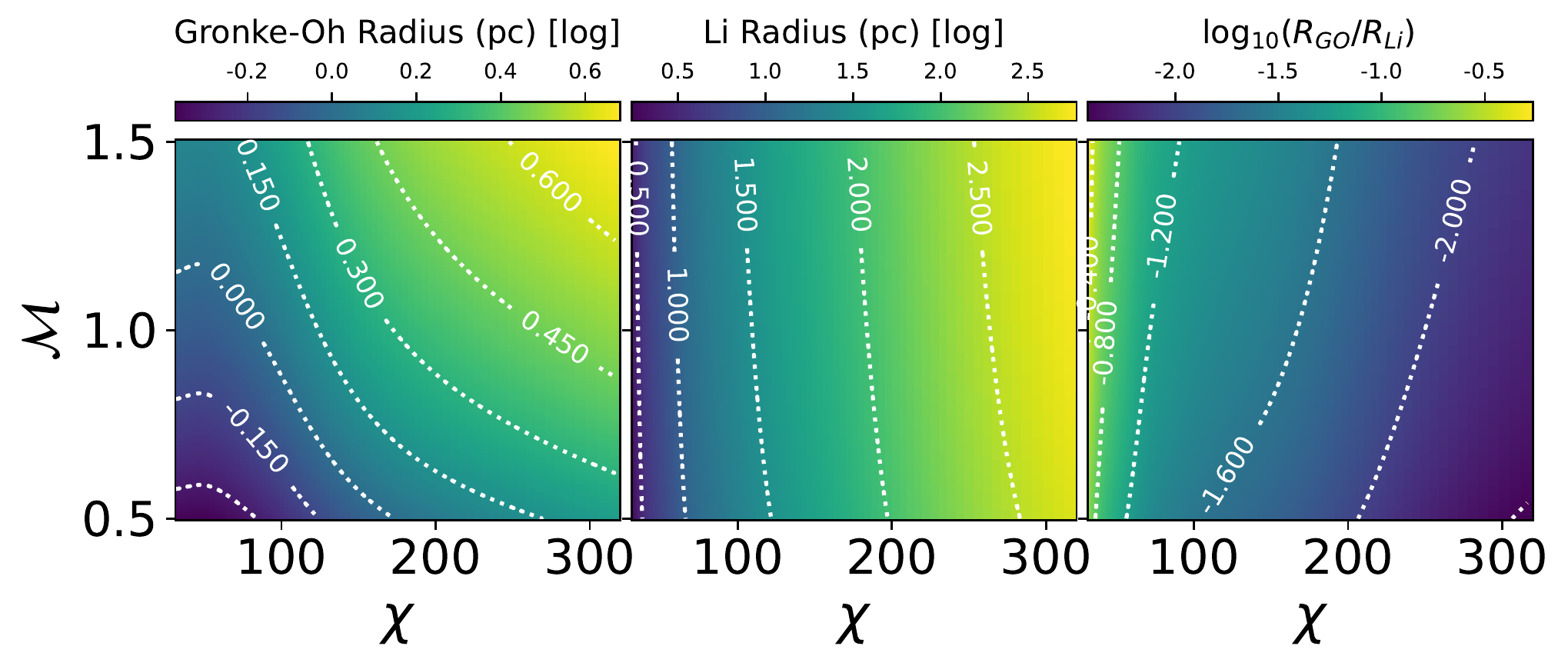}
\end{subfigure}
\caption{The Gronke-Oh radius ($\log$; Eq.~\ref{eq:GO_radius}; left panel), the Li radius ($\log$; Eq.~\ref{eq:Li_radius}; middle panel), and their ratio (right panel) as a function of the density contrast $\chi$ and the Mach number $\mathcal{M}$. The variation of the two radii as a function of $\chi$ and $\mathcal{M}$ is different qualitatively and qauntitatively. We vary these two parameters ($\chi$, $\mathcal{M}$) in our cloud-crushing simulations with cooling to test the two criteria for the growth of cold/dense gas. For most of the relevant parameter regime, the Li radius exceeds the Gronke-Oh radius by more than a factor of 10, with the discrepancy increasing with $\chi$.}
\label{fig:GOLidep}
\end{figure*}

\label{sec:analytic}
According to \citetalias{gronke2018}, \citetalias{gronke2020} cloud growth occurs whenever the cooling time of the mixed gas (evaluated at $T_{\rm mix} \approx \sqrt{T_{\rm cl}T_{\rm hot}}$; see Table~\ref{tab:Variable_defn} for various parameters of the cloud-crushing problem) is shorter than the cloud-crushing time; i.e., $t_{\rm cool,mix}/t_{\rm cc}< 1$. In terms of the initial cloud radius, this criterion corresponds to a cloud size larger than
\begin{equation}
    R_{\rm GO} \approx 2 {\rm~pc} \frac{T_{\rm cl,4}^{\frac{5}{2}}\mathcal{M}}{P_{3}\Lambda _{\rm mix,-21.4}}\frac{\chi }{100} = 2 {\rm~pc} \frac{T_{\rm cl,4}^{\frac{3}{2}}\mathcal{M}}{n_{\rm cl,0.1}\Lambda _{\rm mix,-21.4}}\frac{\chi }{100},
	\label{eq:GO_radius}
\end{equation}
where $T_{\rm cl,4}\equiv  ( T_{\rm cl}/10^{4} {\rm K} )$, $P_{3}\equiv nT/(10^{3}{\rm cm}^{-3} {\rm K})$, $\Lambda _{\rm mix,-21.4}\equiv \Lambda (T_{\rm mix})/(10^{-21.4}{\rm erg~cm}^{3} {\rm s}^{-1})$, and
$n_{\rm cl,0.1}\equiv (n_{\rm cl}/0.1{\rm~cm}^{-3})$. Recall that the cooling time is
\begin{equation}
t_{\rm cool} \equiv \frac{3}{2} \frac{k_{B}T}{n\Lambda \left( T \right)}, 
\label{eq:tcool}
\end{equation}
where $\Lambda(T)$ is the cooling function such that the energy loss due to radiative cooling rate per unit volume for a given number density and temperature is given by $n^2 \Lambda(T)$.

In contrast, \citetalias{Li2020} (and also \citealt{Sparre2020}) claim that cloud growth due to radiative cooling occurs when $t_{\rm cool,hot}/t_{\rm life,pred} =  t_{\rm cool,hot}/(10t_{\rm cc}\overline{f})< 1$ (see section 3.4 and Eq. 19 in \citetalias{Li2020}). The radius according to this criterion for cold gas growth, expressed in terms of the same parameters as Eq.~\ref{eq:GO_radius}, should exceed (using Eqs.~\ref{eq:tlife_pred},~\ref{eq:fbar})
\begin{equation}
    R_{\rm Li} \approx 15.4 {\rm~pc} \frac{T_{\rm cl,4}^{\frac{12}{13}}{\mathcal{M}^{\frac{4}{13}}}}{n_{\rm cl,0.1}\Lambda _{\rm hot,-21.4}^{\frac{10}{13}}}\left ( \frac{\chi }{100} \right )^{\frac{20}{13}},
	\label{eq:Li_radius}
\end{equation}
where $\Lambda _{\rm hot,-21.4}\equiv \Lambda (T_{\rm hot})/(10^{-21.4}{\rm erg~cm}^{3} {\rm s}^{-1})$. Note that the cooling function in Eqs.~\ref{eq:GO_radius} \&~\ref{eq:Li_radius} are evaluated at $T_{\rm mix}$ and $T_{\rm hot}$, respectively. For typical values of the parameters, $T_{\rm cl}=10^{4}K$, $\mathcal{M}=1.0$, $\chi =100$, and $n_{\rm cl}=0.1{\rm~cm}^{-3}$, the corresponding Gronke-Oh and Li radii are approximately 1.1 pc and 26.5 pc, respectively. Thus, there exists a major discrepancy between the two criteria. Figure~\ref{fig:GOLidep} shows the Gronke-Oh and Li radii, and their ratio as a function of the density contrast ($\chi$) and the Mach number ($\mathcal{M}$) for a fixed cloud temperature $T_{\rm cl}=10^4$ K and density $n_{\rm cl}=0.1$ cm$^{-3}$. By comparing Eqs.~\ref{eq:GO_radius} \&~\ref{eq:Li_radius}, note that the ratio is independent of the cloud density but depends on the other parameters.

\begin{figure}
	\includegraphics[width=\columnwidth]{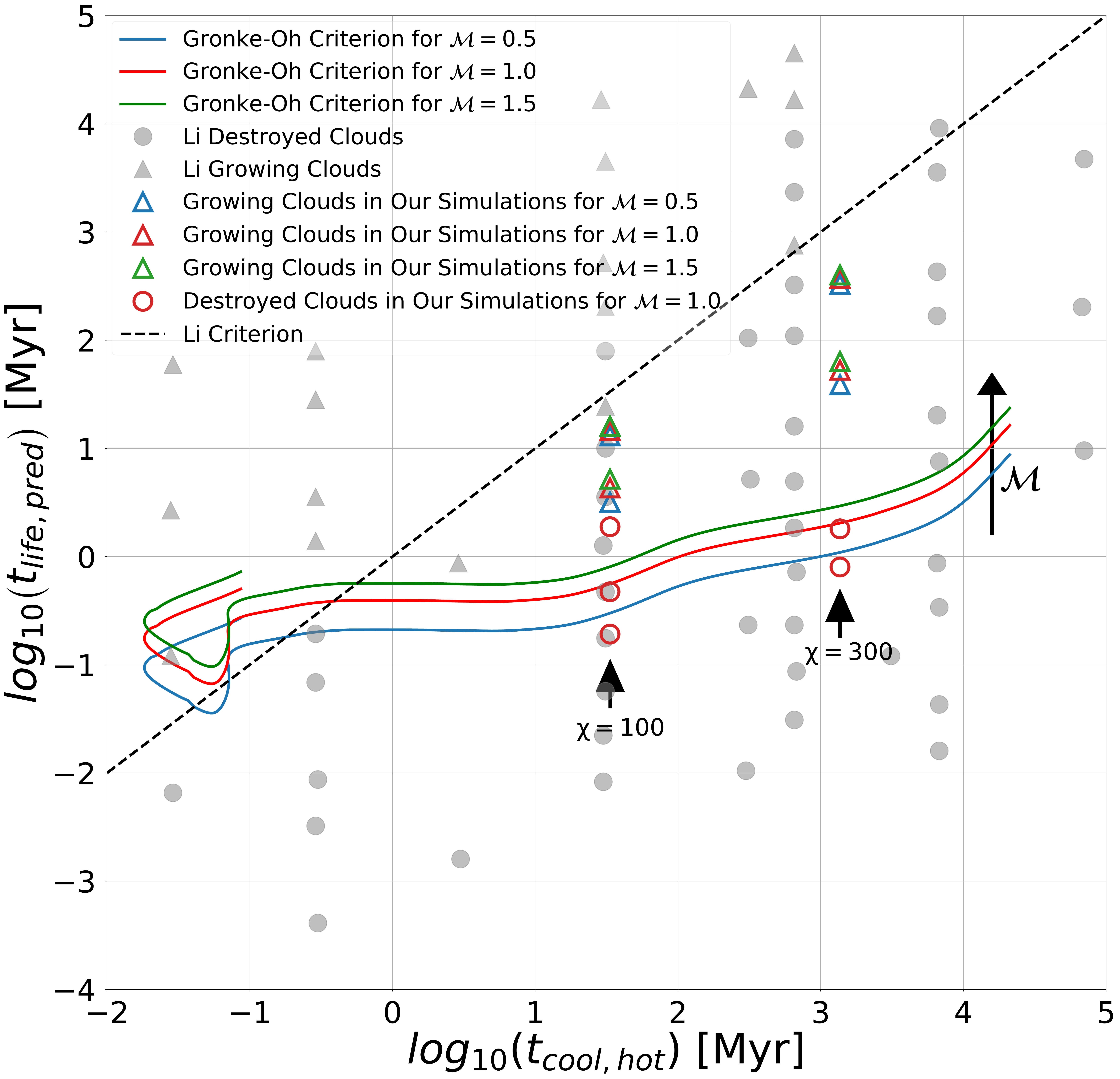}
    \caption{The results of our cloud-crushing simulations (colored open symbols) superimposed on the points taken from the bottom panel of Figure 3 in \citetalias{Li2020} (grey filled symbols). Circles denote the runs where the dense gas is ultimately destroyed and triangles where it grows. The dashed line shows the Li criterion for the growth of clouds in the cloud-crushing problem and the three colored solid lines show the Gronke-Oh criterion for different Mach numbers. Our simulation symbols are colored according to the Mach number of these lines. We find cloud growth in the region between the Li and the Gronke-Oh thresholds, and cloud destruction occurs below the Gronke-Oh threshold. Notice the red open circle lying above the Gronke-Oh radius for $\mathcal{M}=1$ (solid red line), indicating a cloud larger than $R_{\rm GO}$ showing dense mass destruction and hinting that more simulations are needed to precisely map out the threshold $t_{\rm cool, mix}/t_{\rm cc}$.    Our simulations shown in this plot correspond to a density contrast of either $\chi=100$ or $\chi=300$. All our simulations and the three solid lines use $n_{\rm cl}=0.1$ cm$^{-3}$ and $T_{\rm cl}=10^4$ K (these line move down/up with an increasing/decreasing $n_{\rm cl}$).
    }
    \label{fig:Overlay}
\end{figure}

Figure~\ref{fig:Overlay} shows a comparison of the Gronke-Oh and Li criteria for the growth of cold gas in the cloud-crushing problem, expressed in terms of the key parameters of \citetalias{Li2020}, $t_{\rm cool, hot}$ (the cooling time of the hot wind) and $t_{\rm life, pred}$ (the predicted lifetime in the cloud-crushing problem; see Eqs.~\ref{eq:tlife_pred} \&~\ref{eq:fbar}).\footnote{We use $t_{\rm life,pred}-t_{\rm cool, hot}$ coordinates suggested by \citetalias{Li2020} for this comparison because we were not readily able to get their individual simulation parameters ($\chi$, $\mathcal{M}$, $n_{\rm cl}$, $T_{\rm cl}$) and we did not want to misquote their results.} In this space, the Li criterion is simply given by a line of slope unity passing through the origin. Notice that the cooling times $t_{\rm cool, mix}$ and $t_{\rm cool, hot}$, in terms of the cloud-crushing parameters ($n_{\rm cl}$, $T_{\rm cl}$, $\mathcal{M}$, $\chi$, $R_{\rm cl}$), depend on $n_{\rm cl}$, $T_{\rm cl}$ and $\chi$, cloud-crushing time $t_{\rm cc}$ on $R_{\rm cl}$, $\mathcal{M}$ and $T_{\rm cl}$, and predicted lifetime $t_{\rm life,pred}$ on all of them except $\chi$. The three colored solid lines in Figure~\ref{fig:Overlay} show the Gronke-Oh criterion in terms of $t_{\rm cool,hot}$ and $t_{\rm life, pred}$ for three different Mach numbers, and for $n_{\rm cl}=0.1$ cm$^{-3}$ and $T_{\rm cl}=10^4$ K. For each of these curves, we fix $\mathcal{M}$, $n_{\rm cl}$ and $T_{\rm cl}$ and vary $\chi$. For each $\chi$ and $\mathcal{M}$, we calculate the cooling time of the hot phase $t_{\rm cool,hot}$ and the threshold Gronke-Oh radius (Eq. \ref{eq:GO_radius}), and then the predicted cloud lifetime $t_{\rm life,pred}$ (Eqs. \ref{eq:tlife_pred}, \ref{eq:fbar}) is calculated with the cloud radius set to the Gronke-Oh radius.
Note that at small values of $t_{\rm cool, hot}$, different values of $\chi$ can give the same $t_{\rm cool, hot}$\footnote{This happens below $\chi \approx 2$, where the cooling function is locally steeper than $T^2$.} but different $R_{\rm GO}$ and therefore $t_{\rm life,pred}$. The Gronke-Oh and Li criteria are both qualitatively and quantitatively very different, and their difference increases with an increasing $\chi$ (consistent with the right panel of Figure~\ref{fig:GOLidep}).

\section{Numerical Simulations}
\label{sec:num}

Now that we have highlighted the large discrepancy between the analytic Gronke-Oh and Li criteria for cold cloud growth in the cloud-crushing problem with cooling, we turn to numerical simulations to check whether they agree with any of these criteria.

We set up three-dimensional hydrodynamical simulations in a uniform ($\Delta x=\Delta y =\Delta z = d_{\rm cell}$) Cartesian grid corresponding to the classical cloud-crushing problem along with optically thin radiative cooling of collisionally ionized gas similar to \citetalias{gronke2018}, \citetalias{gronke2020}. Since the physical set up is well-known, we do not repeat all the details here and just present the key results. We implement our numerical experiments in PLUTO 4.3, a conservative Godunov hydrodynamics code (\citealt{PLUTO}). We place a stationary, dense, spherical cloud of cold gas of radius $R_{\rm cl}$ at a temperature of $T_{\rm cl}=10^{4}$ K (note that \citetalias{gronke2018}, \citetalias{gronke2020} assume a slightly higher  cloud temperature of $4\times 10^{4}$ K) and number density $n_{\rm cl}=0.1{\rm~cm}^{-3}$ in our simulation domain and impose a steady hot wind outside it. All the boundaries are outflowing except the one in the upstream direction where we impose a constant inflow of the hot gas at a fixed density, temperature and velocity (as measured in the lab frame). We also try periodic boundary conditions in the transverse direction to check the robustness of our results (see Appendix \ref{app:box-size}).

We use a cloud-tracking scheme which continuously changes the instantaneous frame of reference to the center of mass of the cloud material (tagged by a passive scalar) so that we can follow the cloud/mixed gas and prevent it from quickly moving out of the box (e.g., \citealt{Shin2008,Dutta_2019}). Nevertheless, we still we use fairly long boxes to prevent the cloud-wind interaction region from moving out of the computational domain. The typical box-size for our simulations is 400 $R_{\rm cl}$ along the wind direction and 30 $R_{\rm cl}$ in the orthogonal directions. This box is large enough such that the passive scalar (marking the initial cloud) does not leak out of the boundaries (this is ensured by monitoring the volume integral of the passive scalar). We primarily focus on simulations with a resolution of $R_{\rm cl}/d_{\rm~cell}=8$, but we also perform simulations up to $R_{\rm cl}/d_{\rm~cell}=64$ (this high resolution run has a smaller box-size $30 R_{\rm cl} \times 15 R_{\rm cl} \times 15 R_{\rm cl}$). Appendices~\ref{app:box-size} \&~\ref{app:res} study the impact of the box-size and resolution, respectively.

We use a tabulated cooling function generated by CLOUDY (\citealt{2017RMxAA..53..385F}) for a solar metallicity plasma (\citealt{2009ARA&A..47..481A}) assuming collisional ionization equilibrium (CIE). This cooling table is available with the publicly available version of the PLUTO code and cooling is implemented using ODE solvers (allowing for stiff cooling) described in \citet{Tesileanu2008}. We set the cooling function to zero below $T_{\rm cl}$ to crudely mimic heating due to the UV background (note that some runs in \citetalias{Li2020} use self-shielding that allows gas to cool to $\rm \sim 10K$). We assume that the gas is fully ionized and the mean molecular weight is constant for temperatures $\rm \gtrsim 10^4 K$. In Table~\ref{tab:numerical-solver} we list the various physical and numerical options implemented in our simulations. Most of our simulations are run for a timescale shorter than the cooling time of the hot wind because we wish to study the cooling of the mixed gas produced in the boundary layer. For the CGM one also has to worry about the thermodynamics of the diffuse hot phase, which is beyond the scope of the present work.

\begin{table}
	\begin{center}
	\caption{Various code parameters$^\dag$}
	\label{tab:numerical-solver}
	\begin{tabular}{lr}
		\hline
		\hline
		Geometry & Cartesian\\
		Solver & HLLC \\
		Cooling & Tabulated function (solar metallicity)\\
		Flux form & Dimensionally-split\\
		External forces & Absent\\
		Thermal Conduction & Absent\\
		Self-gravity & Absent\\
		Reconstruction & Linear TVD \\
		Time stepping & RK2\\
		Viscosity & Non-explicit code viscosity\\
		Equation of state & Ideal gas, $\gamma=5/3$\\
		CFL limiting value & 0.2\\
		\hline
	\end{tabular} 
	\end{center}
	$^\dag$ see the \href{http://plutocode.ph.unito.it}{PLUTO code manual} (v4.3) for details.
\end{table}

We initialize the setup in pressure equilibrium. The Mach number of the background wind is chosen from  $\mathcal{M} = \left \{0.5,1.0,1.5\right \}$ and the density contrast from $\chi=\left \{100,300\right \}$ to cover a considerably wide range in the parameter space relevant to the CGM. For a particular combination of $\mathcal{M}$ and $\chi$ (and our fixed $n_{\rm cl}$ and $T_{\rm cl}$), we obtain the corresponding Gronke-Oh and Li radii, $R_{\rm GO}$ and $R_{\rm Li}$, from Eq.~\ref{eq:GO_radius} and Eq.~\ref{eq:Li_radius} respectively. Since numerical experiments with cloud sizes between the Gronke-Oh and Li radii are essential to resolve the discrepancy between the two criteria, we parameterize the cloud size as  
\begin{equation}
\label{eq:zeta}
    R_{\rm cl}= R_{\rm Li}^{\zeta }R_{\rm GO}^{1-\zeta},
\end{equation}
where we choose the "interpolation parameter" $\zeta$ between 0.3 and 1 for the runs with $R_{\rm GO} \leq R_{\rm cl} \leq R_{\rm Li}$. The cloud size between the Gronke-Oh and Li radii correspond to $0<\zeta < 1$, with a smaller $\zeta$ corresponding a smaller size closer to the Gronke-Oh radius (which is always smaller than the Li radius for parameters of interest). We also carry out a few runs with clouds smaller than the Gronke-Oh radius and verify that the clouds smaller than this are indeed destroyed. Table~\ref{tab:initial_radius} lists the various parameters -- $\chi$, $\mathcal{M}$, $R_{\rm cl}$ and $\zeta$ -- for our different simulations.

\begin{table*}
	\begin{center}
	\caption{Physical \& numerical parameters of our different runs}
	\label{tab:initial_radius}
	\begin{tabular}{ccccccccccc} 
		\hline
		$\chi$ & $\mathcal{M}$  & $R_{\rm cl}$ & $\zeta$  & Cooling & $t_{\rm cc}$ & $t_{\rm cool,mix}$ & $R_{\rm cl}/d_{\rm cell}$ & box-size in  &  Boundary &  Cloud  \\
		 &   & (pc) & (Eq.~\ref{eq:zeta})  &   &  (Myr) & (Myr) &  & $R_{\rm cl}$ (L,T,T)  & Condition (T) &  Growth \\
		\hline
		100$^\dag$ & 1.0 &  14.0 & 0.8 & yes & 0.96 & 0.09 &  8,16,32 & (400,30,30) & outflow &  yes \\
		100$^\dag$ & 1.0 &  14.0 & 0.8 & yes & 0.96 & 0.09 & 64 & (30,15,15) & outflow  & yes \\
		100 & 1.0 &  5.47 & 0.5 & yes & 0.37 & 0.09 & 8 & (400,30,30) & outflow  & yes\\
		100 & 1.0 &  2.90 & 0.3 & yes & 0.19 & 0.09 & 8 & (400,30,30) & outflow  & no\\
		100 & 1.0 &  1.0 & -$^\ddag$ & yes & 0.07 & 0.09 & 8 & (400,30,30) & outflow  & no \\
		100 & 1.0 &  0.5 & - & yes & 0.03 & 0.09 & 8 & (400,30,30) & outflow  & no \\
		100 & 1.0 &  14.0 & 0.8 & no & 0.96 & - & 8 & (400,30,30) & outflow  & no \\
		100 & 1.0 &  5.47 & 0.5 & no & 0.37 & - & 8 & (400,30,30) & outflow  & no \\
		100 & 1.5 &  17.0 & 0.8 & yes & 0.78 & 0.09 & 8 & (400,30,30) & outflow  & yes \\
		100 & 1.5 &  7.16 & 0.5 & yes & 0.33 & 0.09 & 8 & (400,30,30) & outflow  & yes \\
		100 & 0.5 &  10.36 & 0.8 & yes & 1.42 & 0.09 & 8 & (400,30,30) & outflow  & yes \\
		100 & 0.5 &  3.49 & 0.5 & yes & 0.48 & 0.09 & 8 & (400,30,30) & outflow  & yes \\
		300 & 1.0 &  169.02 & 0.8 & yes & 11.58 & 0.25 & 8 & (400,30,30) & outflow  & yes \\
		300 & 1.0 &  37.64 & 0.5 & yes & 2.58 & 0.25 & 8 & (400,30,30) & outflow  & yes \\
		300 & 1.0 &  2.8 & - & yes & 0.19 & 0.25 & 8 & (400,30,30) & outflow & no \\
		300 & 1.0 &  1.5 & - & yes & 0.10 & 0.25 & 8 & (400,30,30) & outflow  & no \\
		300 & 1.5 & 202.45 & 0.8 & yes & 9.24 & 0.25 & 8 & (400,30,30) & outflow & yes\\
		300 & 1.5 &  49.01 & 0.5 & yes & 2.24 & 0.25 & 8 & (400,30,30) & outflow & yes\\
		300 & 0.5 &  124.06 & 0.8 & yes & 16.99 & 0.25 & 8 & (400,30,30) & outflow & yes \\
		300 & 0.5 &  23.92 & 0.5& yes & 3.28 & 0.25 & 8 & (400,30,30) & outflow & yes \\
		50 & 1.0 &  6.79 & - & yes & 0.46 & 0.07 & 8 & (400,30,30) & outflow & no \\	
		100 & 1.0 &  26.5 & 1.0 & yes & 1.82 & 0.09 & 8 & (20,10,10) & periodic & yes\\
		100 & 1.0 &  14.0 & 0.8 & yes & 0.96 & 0.09 & 8 & (400,30,30) & periodic  & yes \\
		100 & 1.0 &  14.0 & 0.8 & yes & 0.96 & 0.09 & 8 & (200,10,10) & periodic & yes \\
		100 & 1.0 &  14.0 & 0.8 & yes & 0.96 & 0.09 & 8 & (100,10,10) & periodic & yes \\
		100 & 1.0 &  14.0 & 0.8 & yes & 0.96 & 0.09 & 8 & (20,10,10) & periodic & yes \\
		100 & 1.0 &  10.28 & 0.7 & yes & 0.70 & 0.09 & 8 & (20,10,10) & periodic & yes\\
		100 & 1.0 &  7.5 & 0.6 & yes & 0.51 & 0.09 & 8 & (20,10,10) & periodic & yes \\
		100 & 1.0 &  6.4 & 0.55 & yes & 0.44 & 0.09 & 8 & (20,10,10) & periodic & no \\
		100 & 1.0 &  5.47 & 0.5 & yes & 0.37 & 0.09  & 8 & (100,10,10) & periodic  & yes \\
		100 & 1.0 &  5.47 & 0.5 & yes & 0.37 & 0.09  & 8 & (20,10,10) & outflow & no \\
		100 & 1.0 &  5.47 & 0.5 & yes & 0.37 & 0.09 & 8 & (20,10,10) & periodic & no \\
		\hline
	\end{tabular}
	\end{center}
	$^\dag$ The fiducial runs (different versions differ by resolution and box-size).
	$^\ddag$ For $R_{\rm cl} < R_{\rm GO}$ (Eq.~\ref{eq:GO_radius}) or $R_{\rm cl}>R_{\rm Li}$ (Eq.~\ref{eq:Li_radius}), we do not specify $\zeta$. \\
	The longitudinal boundary conditions are outflow in the downstream direction and wind parameters are imposed in the upstream direction.
\end{table*}

\section{Results and Discussion}
\label{sec:results}
In this section, we discuss the results of our numerical simulations designed to distinguish between the Gronke-Oh and Li criteria for the growth of cold gas in the cloud-crushing problem with cooling. 

\begin{figure*}
\begin{subfigure}{\textwidth}
  \includegraphics[width=0.32\linewidth]{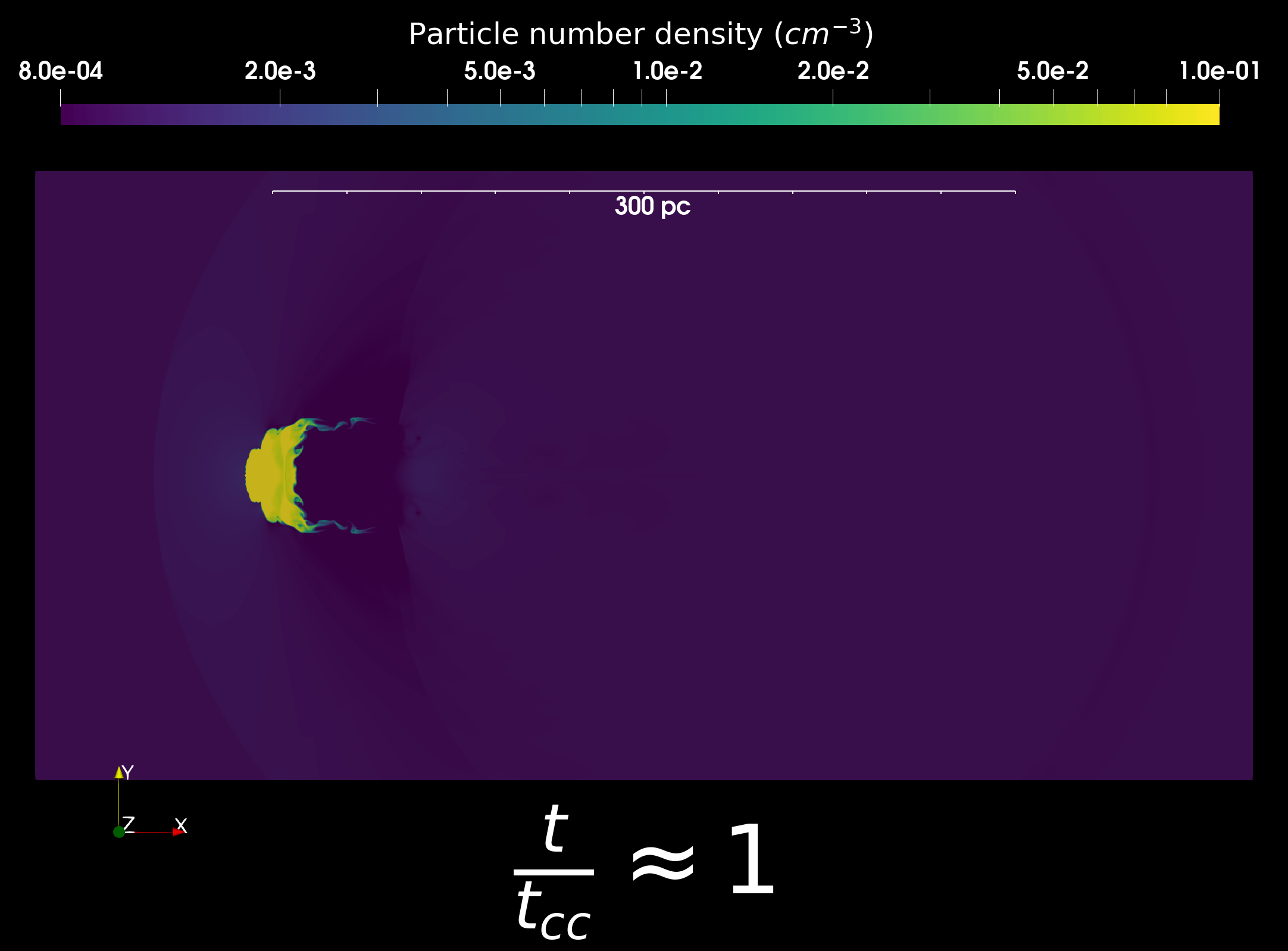}
  \includegraphics[width=0.32\linewidth]{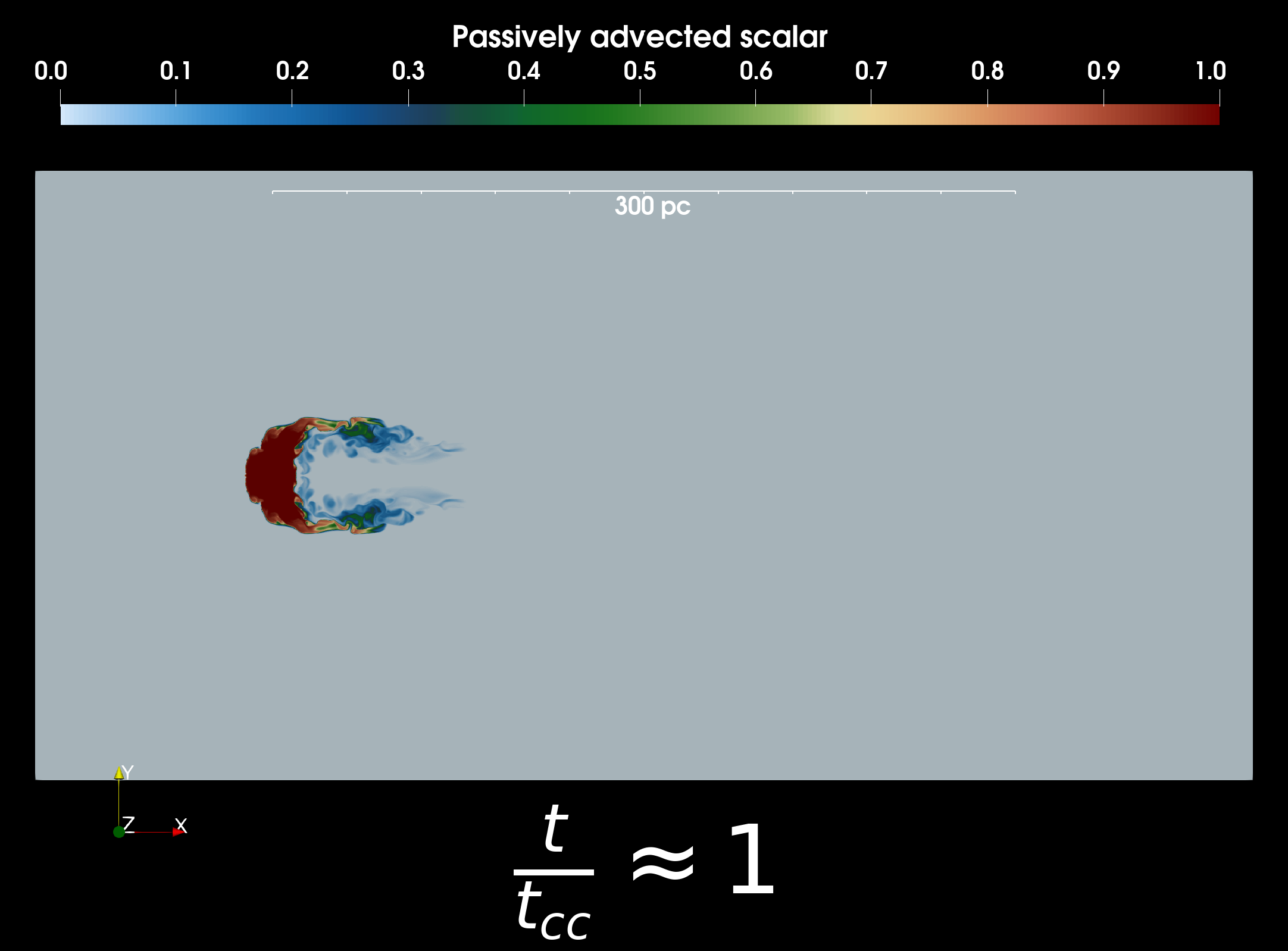}
  \includegraphics[width=0.32\linewidth]{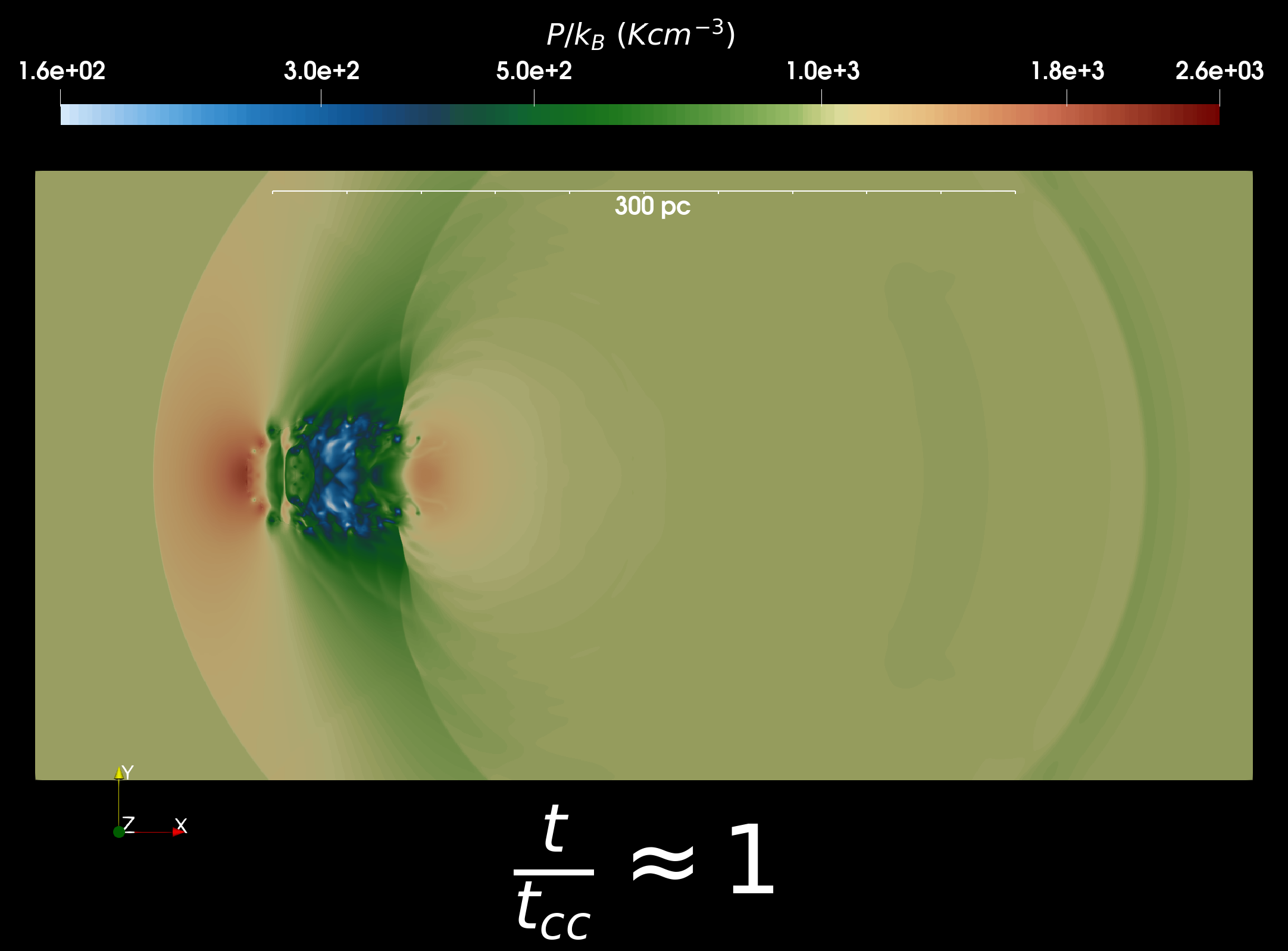}
\end{subfigure}
\begin{subfigure}{\textwidth}
  \includegraphics[width=0.32\linewidth]{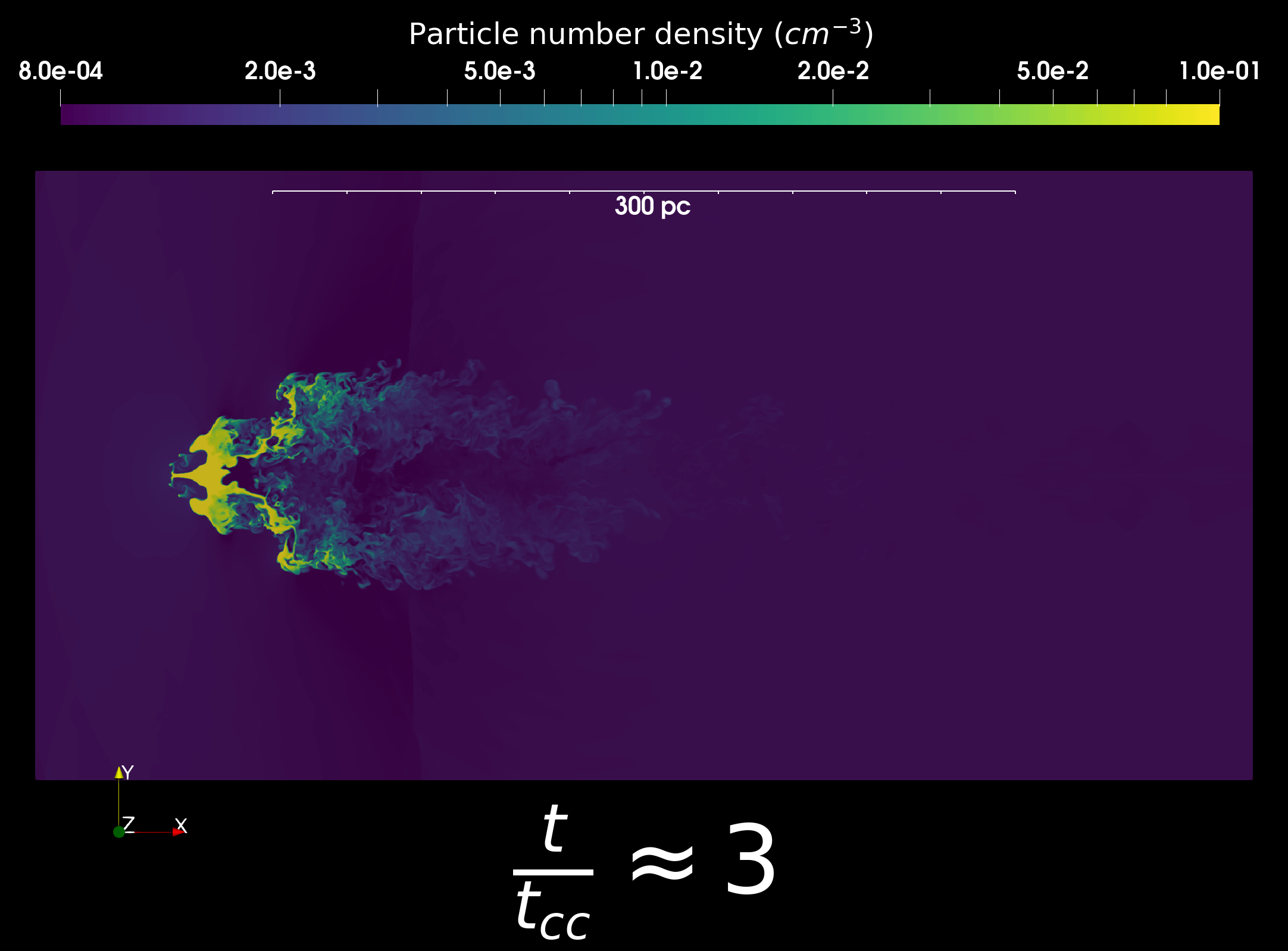}
  \includegraphics[width=0.32\linewidth]{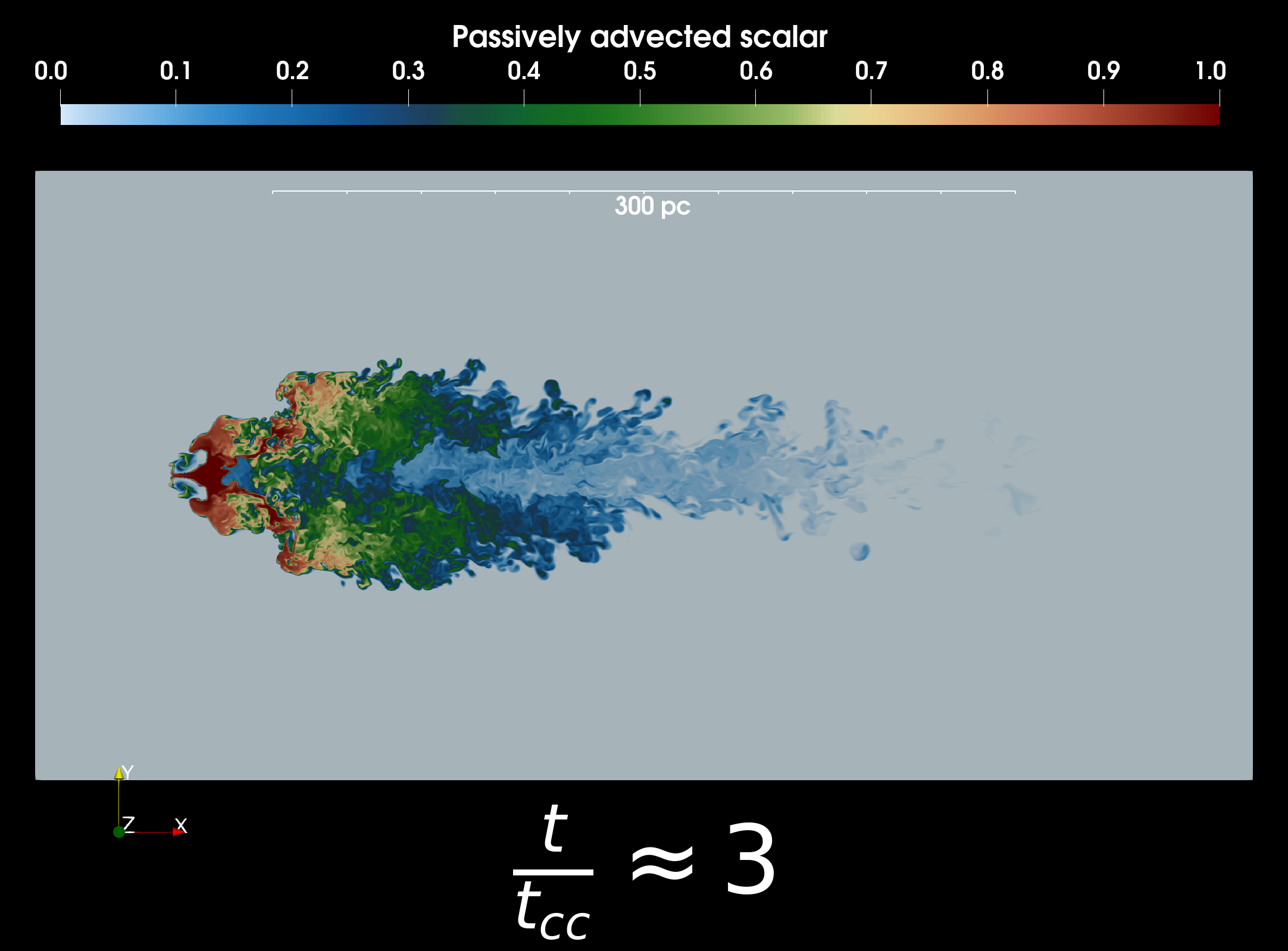}
  \includegraphics[width=0.32\linewidth]{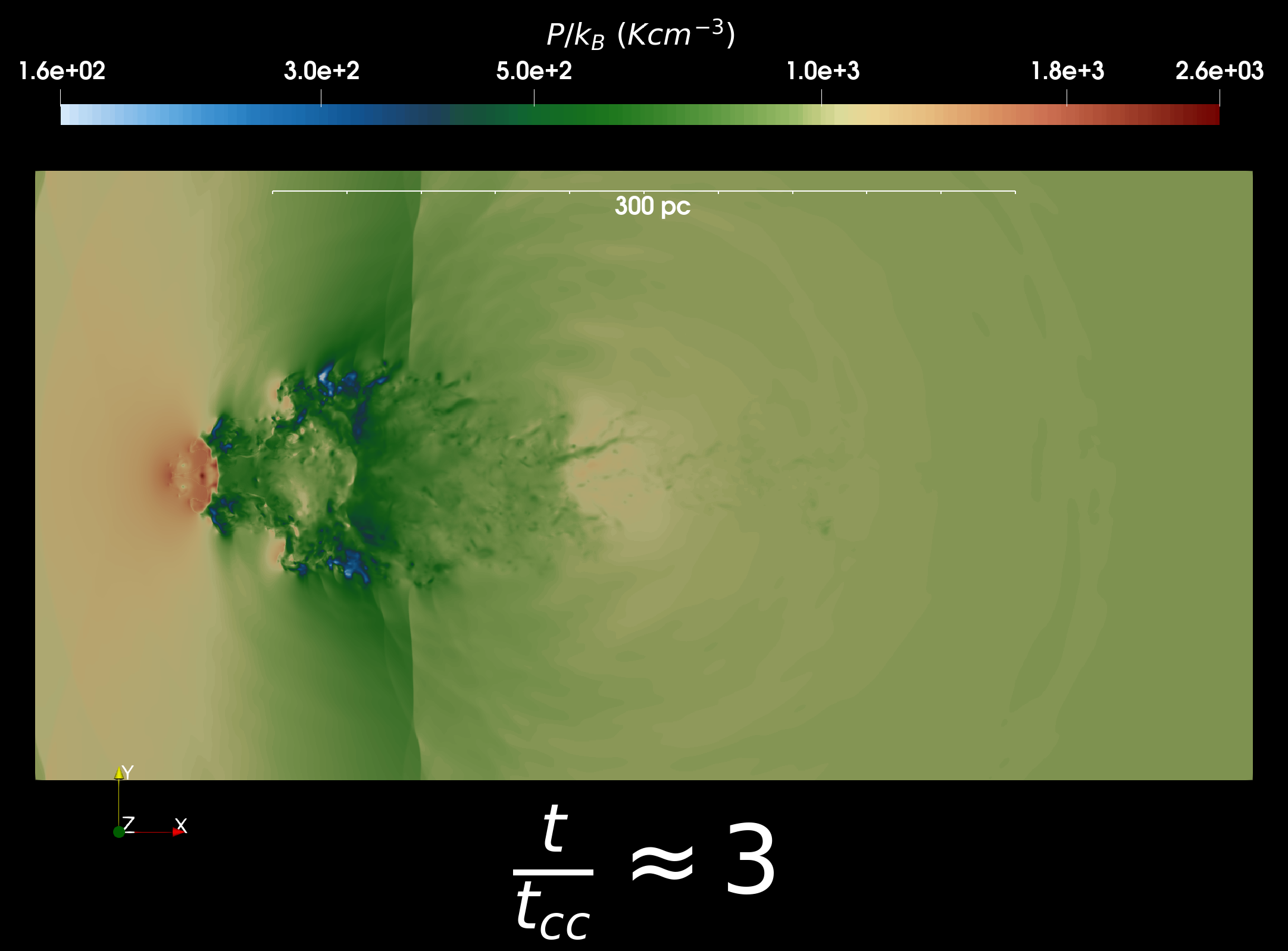}
\end{subfigure}
\begin{subfigure}{\textwidth}
  \includegraphics[width=0.32\linewidth]{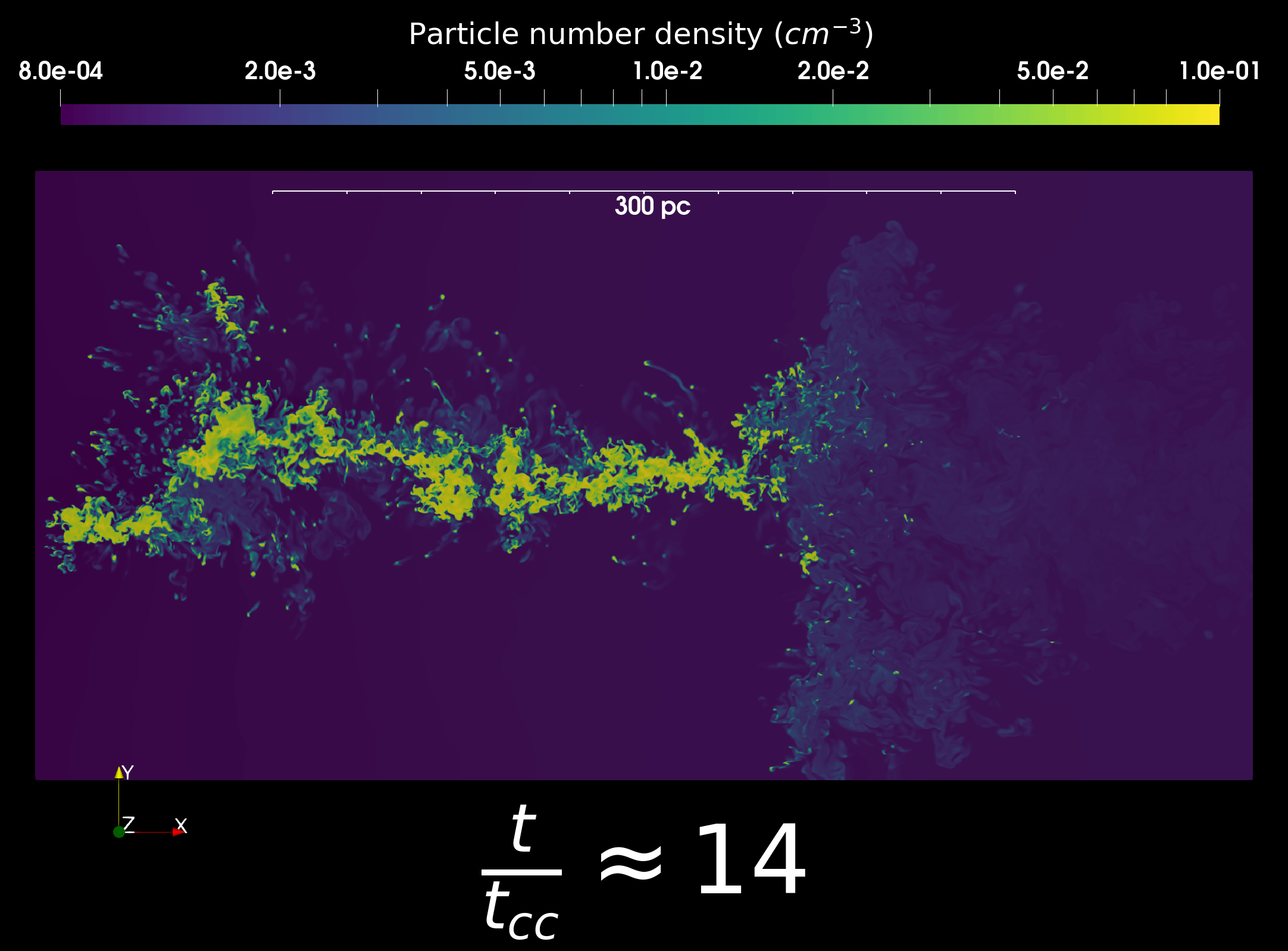}
  \includegraphics[width=0.32\linewidth]{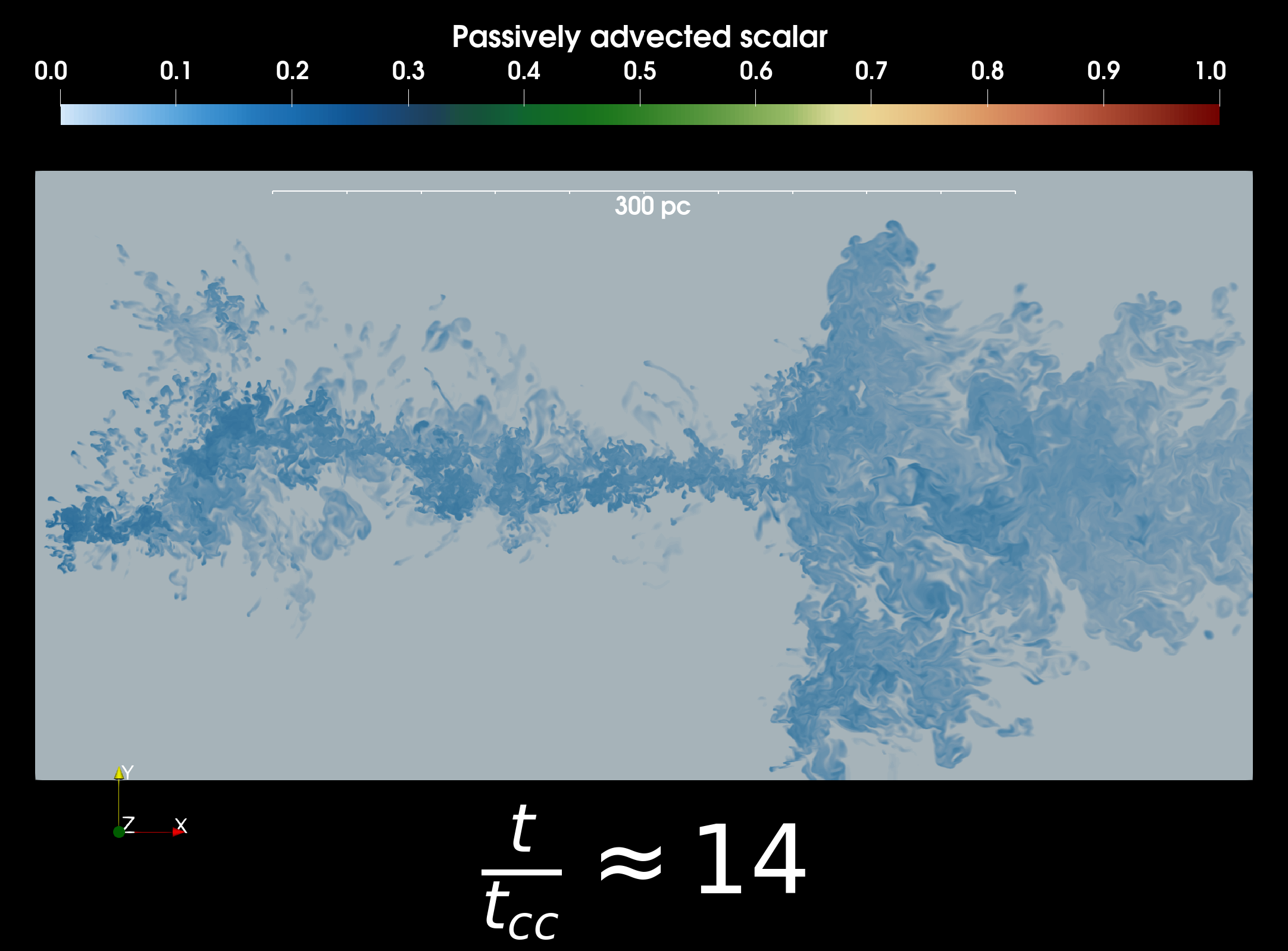}
  \includegraphics[width=0.32\linewidth]{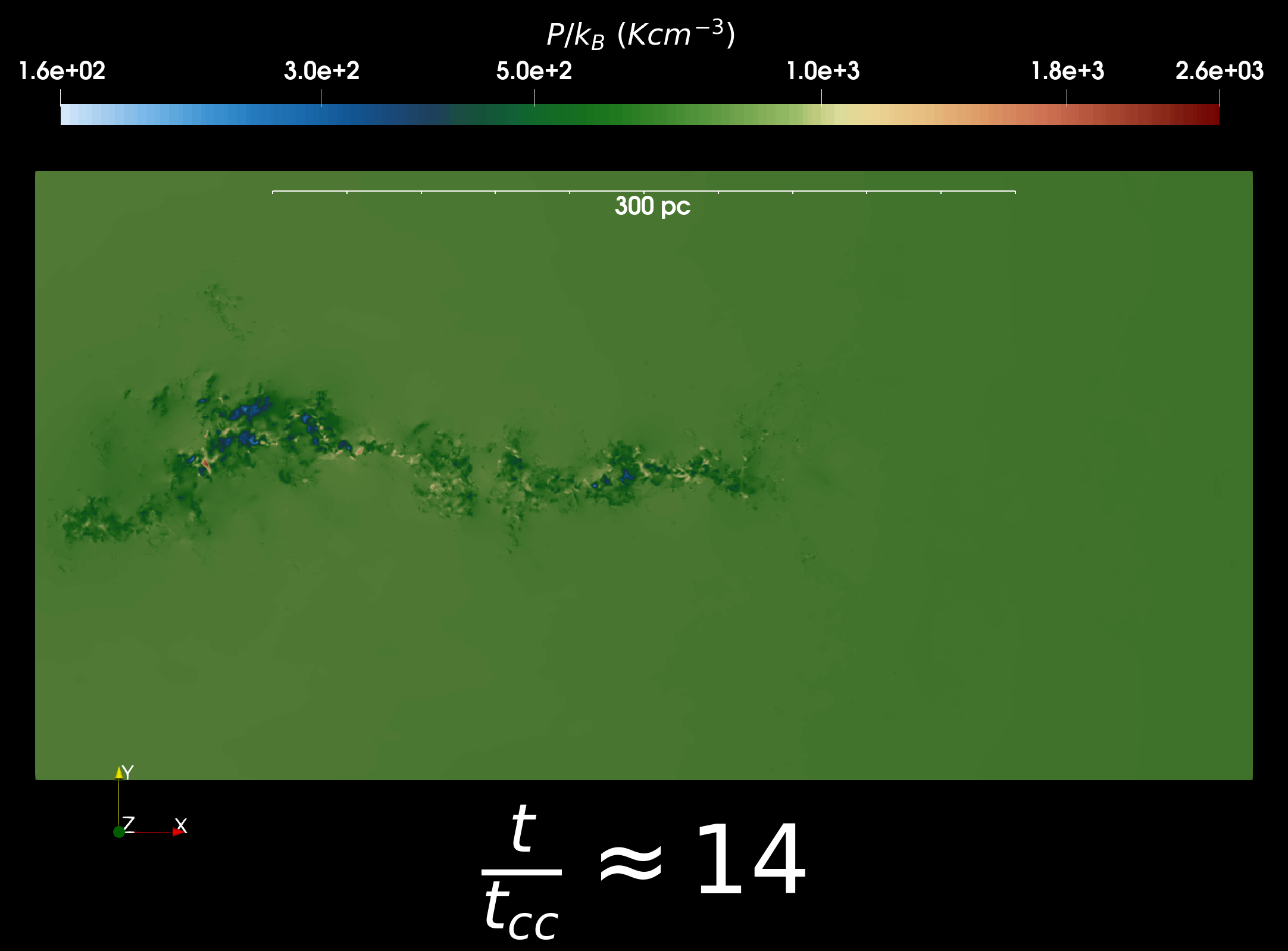}
\end{subfigure}
\caption{The density (left panels), passive scalar (middle panels), and pressure (right panels) slices in the $x-y$ plane through the center at different times for our fiducial high resolution simulation ($R_{\rm cl}=14$ pc, $R_{\rm cl}/d_{\rm cell}=64$; see Table~\ref{tab:initial_radius}). The initial pressure is uniform but a bow shock forms as the wind interacts with the stationary cold cloud. The mixing and cooling in the boundary layer leads to the growth of cold gas and entrainment of the cold gas by the wind at late times. Notice the dense clumps falling on to the filamentary tail at late times and the flaring of the tail at its end. The dense tail has a passive scalar value $\lesssim 0.3$ at $t\approx 14 t_{\rm cc}$, indicating that most of this gas originates in the hot wind.
}
\label{fig:slices}
\end{figure*}

Figure~\ref{fig:slices} shows the mid-plane density (left panels), passive scalar concentration (middle panels), and pressure (right panels) slice plots for the fiducial run with $\chi=100$, $\mathcal{M}=1$ and $R_{\rm cl}=14$ pc, which is predicted to grow according to the Gronke-Oh criterion but not according to the Li criterion. These snapshots are made for the highest resolution run with $R_{\rm cl}/d_{\rm cell}=64$ but a smaller box-size $30 R_{\rm cl} \times 15 R_{\rm cl} \times 15 R_{\rm cl}$ rather than for the standard runs with larger boxes to better visualize the small-scale features. The price we pay is that a part of the turbulent mixing layer moves out of the computational domain. 

With time in Figure~\ref{fig:slices}, we see the formation of a turbulent wake with intermediate densities and intermediate passive scalar concentrations. While pressure is uniform in the initial state, a bow shock is formed ahead of the cloud that becomes weaker with time because of momentum transfer between the wind and the cloud. The bow shock quickly moves out of the computational domain. At late times, the turbulent wake becomes long and filamentary, with the intermediate density clouds cooling and falling on to the tail. The cooler portions of the wake collapse on to the filamentary tail, which has a smaller pressure, in form of a quasi-steady, pressure-driven cooling flow (Dutta et al. in prep.). Such a collapse of the wake is not observed in the simulations with weak cooling in which the cold gas is eventually completely destroyed. Notice the flaring of the turbulent tail at the end, very clearly seen in the passive scalar snapshots at late times. Similar features are observed in our low resolution simulations but obviously not in this great detail. Appendix~\ref{app:fractal} shows that the surfaces with the shortest cooling times are highly corrugated, with a fractal-like structure.

Now that we have looked at the evolution of the morphology of the wake in the cloud-crushing problem with cooling, we study the effects of different parameters on cloud survival and compare our result with the Gronke-Oh and Li survival criteria. In this context, we look at some of the standard volume-integrated diagnostics that quantify the mass growth of the dense/cold gas and momentum transfer (entrainment) between different phases in our simulations. We analyze the statistical properties of the multiphase wake using probability distribution functions (PDFs) of the different quantities and elucidate the physics of the turbulent multiphase wake.

\subsection{Cloud survival \& growth}
\begin{figure*}
\begin{subfigure}{\textwidth}
\includegraphics[width=0.99\linewidth]{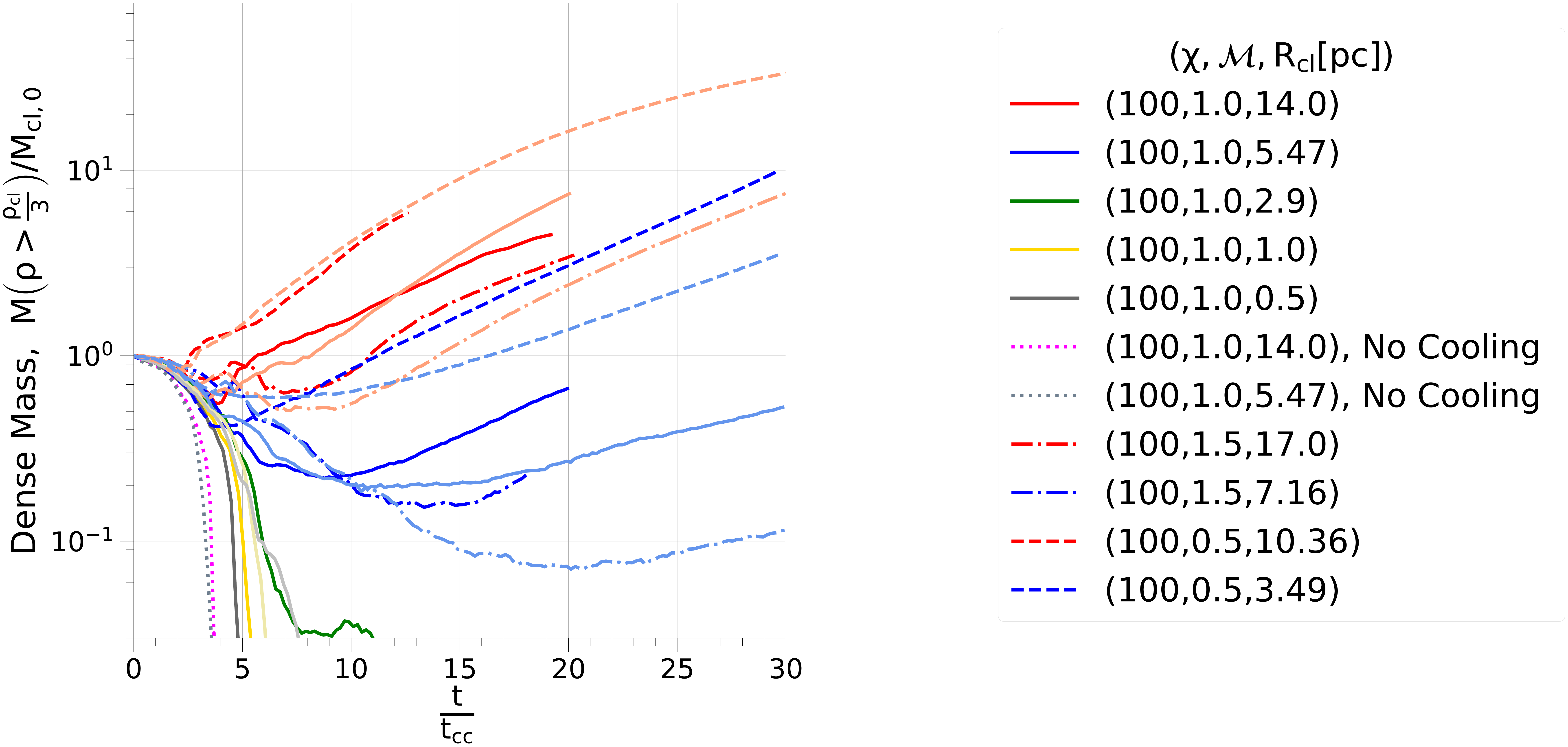}
\end{subfigure}
\begin{subfigure}{\textwidth}
\includegraphics[width=0.90\linewidth]{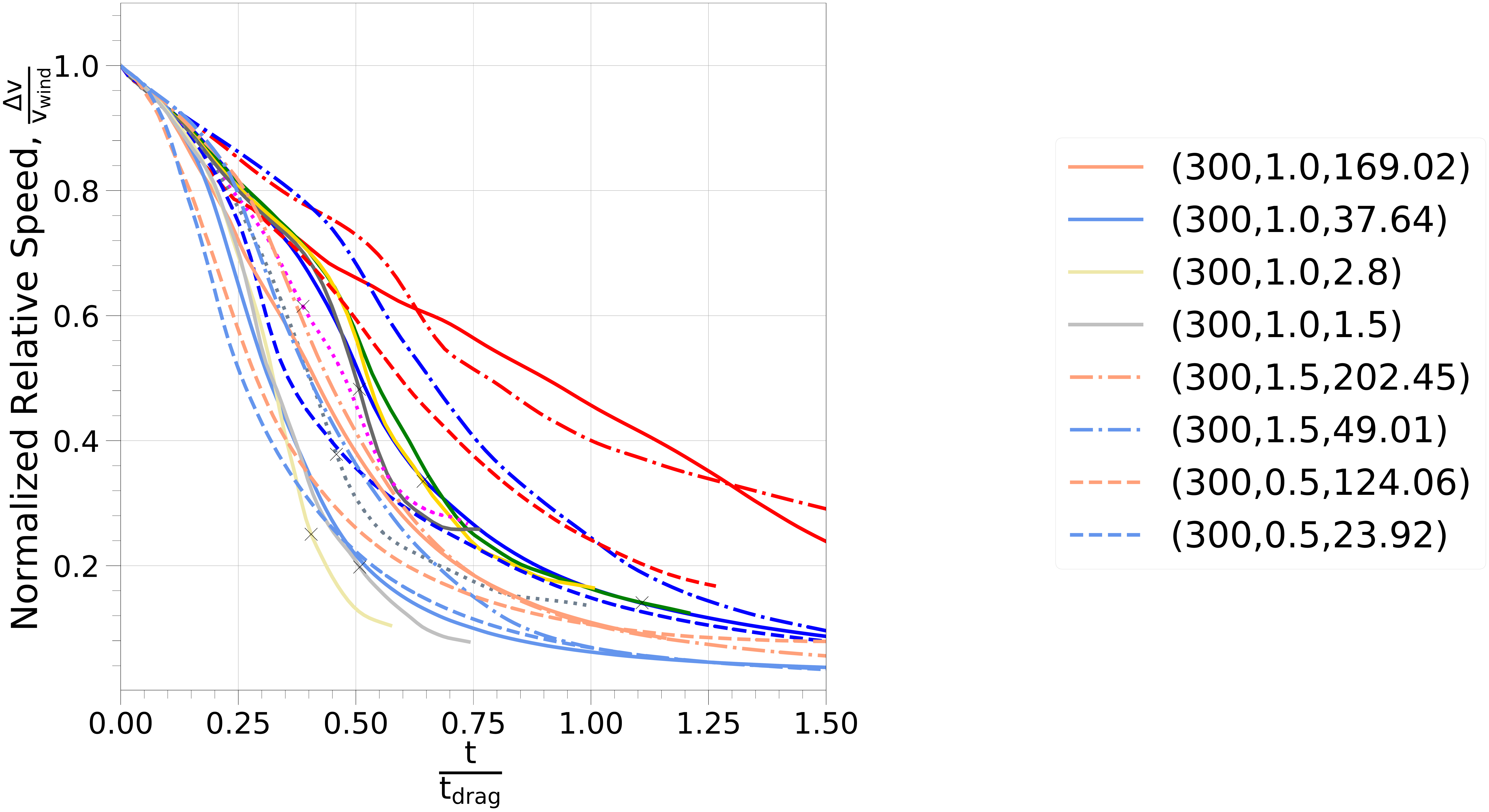}
\end{subfigure}

\caption{\textbf{Top: }The mass evolution of the dense gas $\left ( \rho > \rho _{\rm cl}/3 \right )$ for our runs with various density contrasts ($\chi$), the wind Mach number ($\mathcal{M}$), and the initial cloud radii. The mass is normalized to the initial cloud mass. The runs without cooling show cloud destruction after a few cloud-crushing times (\citealt{klein1994}). Even with radiative cooling, clouds smaller than the Gronke-Oh radius do no show cold mass growth. Larger clouds show an initial dip in the dense mass but it grows later as gas from the hot wind is mixed and cooled into the dense tail (see the bottom panels of Figure~\ref{fig:slices}). The green solid line corresponds to the red circle in Figure~\ref{fig:Overlay} above the red solid line for a cloud larger than the Gronke-Oh radius which is eventually destroyed. \textbf{Bottom:} The speed difference between the wind and the cloud material normalized by the initial wind speed. The black crosses indicate the times when the dense mass gets completely destroyed due to mixing with the hot wind. The decrease in the speed difference indicates momentum transfer due to turbulent mixing between the two phases with an initial velocity difference. Somewhat shallower evolution in some of the runs with cooling may be because of insufficient resolution (see the right panel of Figure~\ref{fig:dme_res}).}
\label{fig:glob-prop-GO}
\end{figure*}

The left panel of Figure~\ref{fig:glob-prop-GO} shows the dense mass (all the mass in computational domain with density $\rho > \rho _{\rm cl}/3$) evolution for all the simulations with $R_{\rm cl}/d_{\rm~cell} = 8$, the box-size (400, 30, 30)$R_{\rm cl}$, and outflow boundary condition in the transverse direction. In absence of cooling, the cold/dense cloud is mixed into the hot/dilute phase in a few cloud-crushing times. The cloud material becomes comoving with the wind on a longer (by $\sim \chi^{1/2}$) drag timescale.

In simulations with cooling, eventually the clouds enter a phase of mass growth if the cloud size is a few times larger than the Gronke-Oh radius (Eq.~\ref{eq:GO_radius}). The simulations with a smaller cloud radius or without cooling eventually show the destruction of dense gas. Since our $0<\zeta<1$ runs (corresponding to a cloud radius smaller than the Li radius but larger than the Gronke-Oh radius) show cloud growth, they are consistent with \citetalias{gronke2018}, \citetalias{gronke2020} and disagree with the Li criterion (Eq.~\ref{eq:Li_radius}).

 If cooling is efficient in the mixed gas formed by shear between the cloud and the wind, the mixed warm gas at the peak of the cooling curve can cool and accrete on to the cold tail and the dense mass can grow. Initially, due to instabilities and mixing, the cloud loses mass, which is clearly observed in the plot till $t\approx 3t_{\rm cc}$ (later for smaller clouds). But for clouds larger than a few $R_{\rm GO}$, the cold gas mass increases at later times because the radiative cooling time of the mixed gas is shorter than the turbulent mixing time ($\sim t_{\rm cc}$). The cold gas mass at $t\approx 18t_{\rm cc}$ for some of the larger clouds is significantly more than the initial cloud mass. In contrast to the cloud-crushing simulations without cooling or  with inefficient cooling, for $R_{\rm cl} \gtrsim R_{\rm GO}$ the mixed warm gas in the boundary layer cools efficiently, creating thermal pressure gradients that enable further inflow of hot gas into the mixing layer. This continuously fed cooling layer entrains the hot wind and leads to the growth of the dense mass. 
 
\subsection{Momentum exchange \& mass entrainment}
\label{sec:mom-ex}
In the cloud-crushing problem in absence of cooling, the timescale for momentum exchange between the cloud and the wind material is $t_{\rm drag} \sim \chi R_{\rm cl}/v_{\rm wind}$, which is longer than the cloud-crushing time by a factor $\sim \chi^{1/2}$. This means that the cloud mixes in the hot wind before it can be pushed substantially by it.\footnote{We note that in absence of cooling, the dense cloud is mixed over 3-4 $t_{\rm cc}$ and the relative velocity between the cloud and wind material (and also dense and diffuse phases; see Appendix \ref{app:deltav_vwind}) halves over $\sim 0.4 t_{\rm drag}$ (Fig. \ref{fig:glob-prop-GO}). Thus the mass and momentum exchange timescales are not very different for $\chi \sim 100$.} The cloud-crushing time scales as the Kelvin-Helmholtz instability growth timescale and the drag timescale corresponds to the time over which the cloud sweeps its own mass in the hot wind. The drag timescale can also be motivated from the Rayleigh drag formula applicable for a turbulent wake; namely the drag force acting on the cold cloud $\sim \rho_{\rm hot} R_{\rm cl}^2 v_{\rm wind}^2$.

The right panel of Figure~\ref{fig:glob-prop-GO} shows the evolution of the relative velocity between the wind and the cloud gas as a function of time for the simulations shown in the top panel. The evolution of the relative velocity provides one measure of momentum transfer between the wind and cloud material. Recall that the initial cloud is traced by a passive scalar $C$, which is set to unity for the cloud and to zero for the wind. Therefore the cloud material velocity at any time is given by
\begin{equation}
\label{}
   {\bf v}_{\rm c} = \frac{\int_V \rho C {\bf v} dV}{\int_V \rho C dV},
\end{equation}
and the wind velocity by
\begin{equation}
\label{}
   {\bf v}_{\rm w} = \frac{\int_V \rho (1-C) {\bf v} dV}{\int_V \rho (1-C) dV},
\end{equation}
where ${\bf v}$ is the fluid velocity, $\rho$ is the mass density, and $V$ denotes the computational volume.\footnote{Note that these velocities are frame-dependent but the relative velocity is not.} We plot the difference between the wind and cloud velocities normalized by the initial wind speed versus time normalized by the classical drag time $t_{\rm drag}$ (see Table~\ref{tab:Variable_defn}). Instead of using passive scalar to track wind and cloud, we can also use temperature to identify them (as done by \citetalias{gronke2018,gronke2020}). We compare various ways of identifying cloud and wind in Appendix~\ref{app:deltav_vwind} and show that the evolution of relative velocity is similar.

Similar to \citetalias{gronke2018}, \citetalias{gronke2020}, we find that entrainment (as measured by the time over which the cloud-wind relative velocity becomes half its initial value) happens on a characteristic timescale $\sim 0.4 t_{\rm drag}$ and there is not much difference between the clouds that grow and the ones that do not. Thus the relative velocity between the wind and the cloud is not much altered even though mass exchange is fundamentally altered due to radiative cooling. Note that for some of the growing clouds, the mass increases by a large factor but the evolution of relative velocity is similar to the destroyed clouds. This must mean that the mixed gas that cools and accretes in cloud tails, leading to the cloud growth, is almost comoving with the cloud (rather than the wind) material (hints for this are also seen in the right panels of Fig. \ref{fig:2d-pdf}). Clearly, more work is needed to understand the detailed momentum transfer between the cloud and the wind.

\subsection{Dependence on density contrast \& Mach number}

In absence of radiative cooling, all clouds are expected to mix into the diffuse hot phase over $\sim 3-4 t_{\rm cc}$. Radiative cooling allows the possibility of the growth of cold/dense mass, provided the cloud radius is large enough. The Gronke-Oh radius (Eq.~\ref{eq:GO_radius}) is larger for a larger density contrast and a higher Mach number. This implies that the clouds with smaller $\chi$ and  $\mathcal{M}$, but a fixed $R_{\rm cl}$, grow faster, as is borne out by the upper panel of Figure~\ref{fig:glob-prop-GO}.

The lower panel of Figure~\ref{fig:glob-prop-GO} shows that the cloud material is close to comoving in all cases after about a drag time, irrespective of whether the cold mass grows or the cold gas is destroyed. There are no significant trends with the density contrast and the Mach number. There may be weak scalings with these parameters but uncovering them is beyond the scope of the present paper. The simulations with faster cooling of the mixed case (equivalently, a larger $R_{\rm cl}$ for fixed $\chi$ and $\mathcal{M}$) show a longer acceleration timescale for the cold cloud but this is likely a numerical artifact due to insufficient resolution (see Appendix~\ref{app:res}).

\subsection{Cloud Dynamics: A detailed look}

The turbulent mixing layer, with gas at intermediate temperature/density/velocity, is an essential element of the dynamics of cloud-crushing both with and without cooling. In absence of cooling, the turbulent boundary layer grows with time and the cold phase is mixed in the hot wind after a few cloud-crushing times. However, the cloud material, which is mixed in the hot wind, takes a longer time ($t_{\rm drag}\sim \chi^{1/2} t_{\rm cc}$; especially for large $\chi$) to become comoving. Thus there is mass transport from the cold to hot phase on $\sim t_{\rm cc}$ timescale and momentum transfer from the hot to cold phase on $\sim t_{\rm drag}$. With cooling, for a sufficiently large cloud, the net mass transfer reverses and is from hot to the cold phase because the mixed layer can cool and accrete on to the cold filamentary tail.

\citet{Begelman1990} consider a mixing layer which entrains mass from the hot and cold phases such that the mean temperature of the mixed phase is 
\begin{equation}
T_{\rm mix} \sim \frac{\dot{M}_{\rm hot} T_{\rm hot} + \dot{M}_{\rm cl} T_{\rm cl}}{\dot{M}_{\rm hot} + \dot{M}_{\rm cl}} \sim \sqrt{T_{\rm hot}T_{\rm cl}},
\end{equation} 
where $\dot{M}_{\rm hot}$ and $\dot{M}_{\rm cl}$ are the mass entrainment rates into the mixed layer from the hot and cold layers, respectively. This estimate assumes that $\dot{M}_{\rm cl}/\dot{M}_{\rm hot} \sim \rho_{\rm cl}v_{\rm in, cl}/(\rho_{\rm hot} v_{\rm in, hot}) \sim v_{\rm in, hot}/v_{\rm in, cl} \sim \chi^{1/2}$ ($v_{\rm in, cl}$ [$v_{\rm in, hot}$] is the inflow velocity from the cold [hot] phase into the mixing layer). These scalings also imply that the mixed layer's longitudinal velocity (like inflow velocity) in the cloud rest frame is smaller (by $\sim \chi^{-1/2}$) compared to the wind velocity.

\citetalias{gronke2018} used the above scalings (which must be refined to match simulations; see \citealt{Fielding_2020,Tan2020}) for a turbulent boundary layer to understand their numerical simulations of cloud-crushing problem with cooling. The cold cloud can grow if the mixed gas, assumed to be at $\sim \sqrt{T_{\rm cl}T_{\rm hot}}$ cools faster than the cloud-crushing time, the timescale over which the cold cloud is mixed into the hot wind in absence of cooling. While this scaling is reasonable for a cooling function that peaks at the intermediate temperatures (which is true for the typical CGM/galactic outflow parameters), it must be refined for different cooling functions.

\subsubsection{1-D Probability distribution functions}

\begin{figure*}
\begin{subfigure}{\textwidth}
  \includegraphics[width=0.33\linewidth]{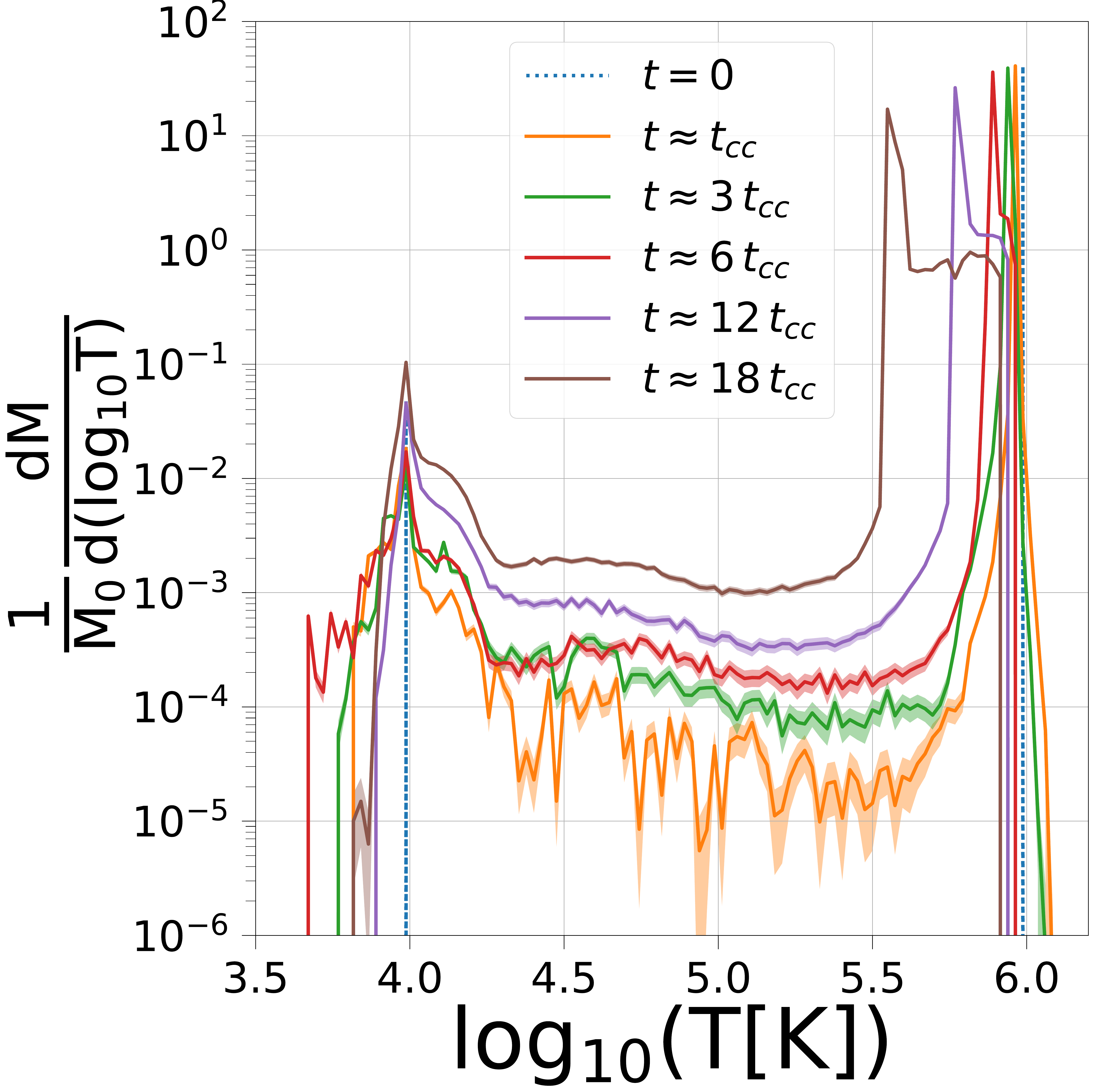}
  \includegraphics[width=0.33\linewidth]{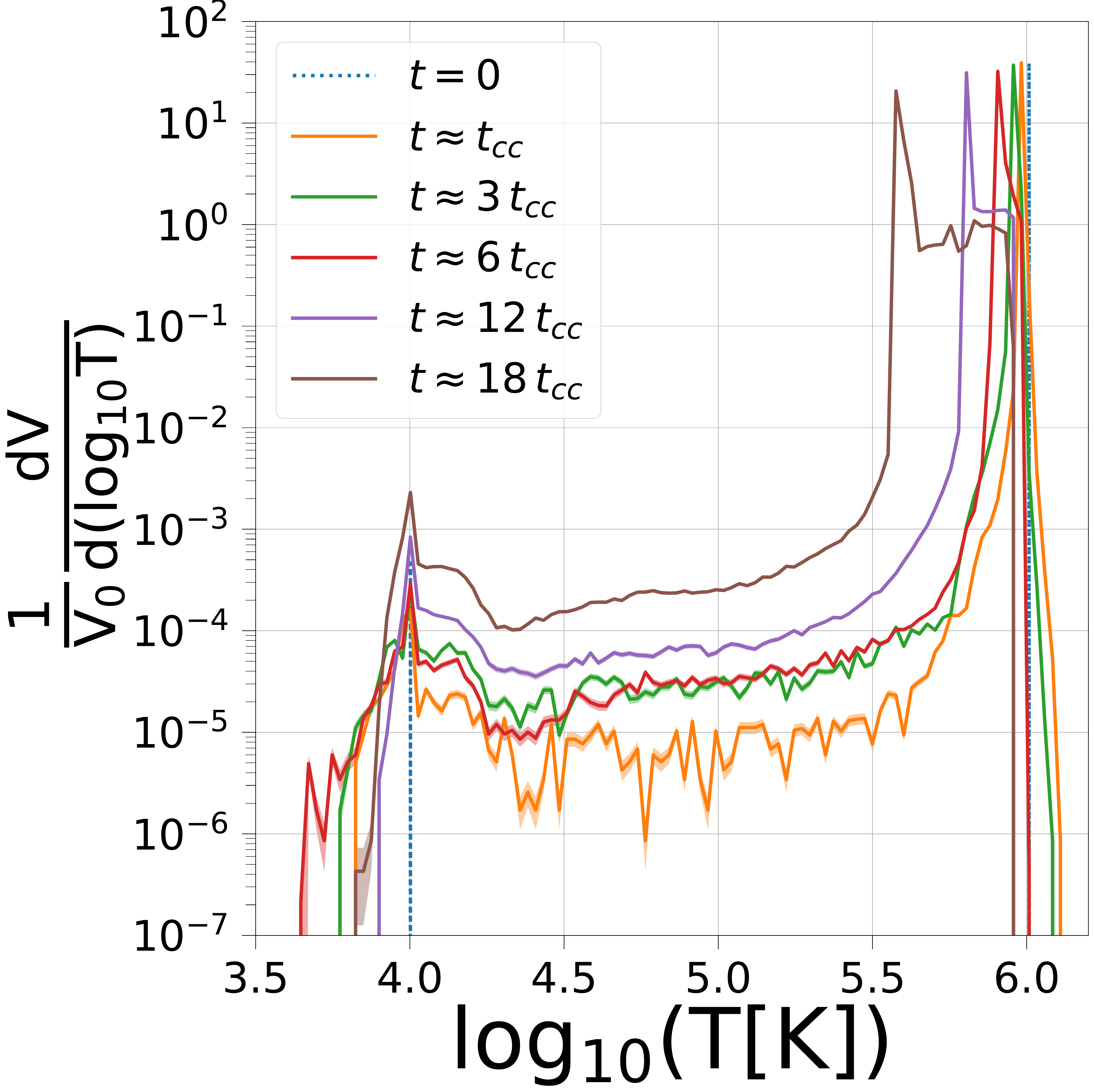}
  \includegraphics[width=0.33\linewidth]{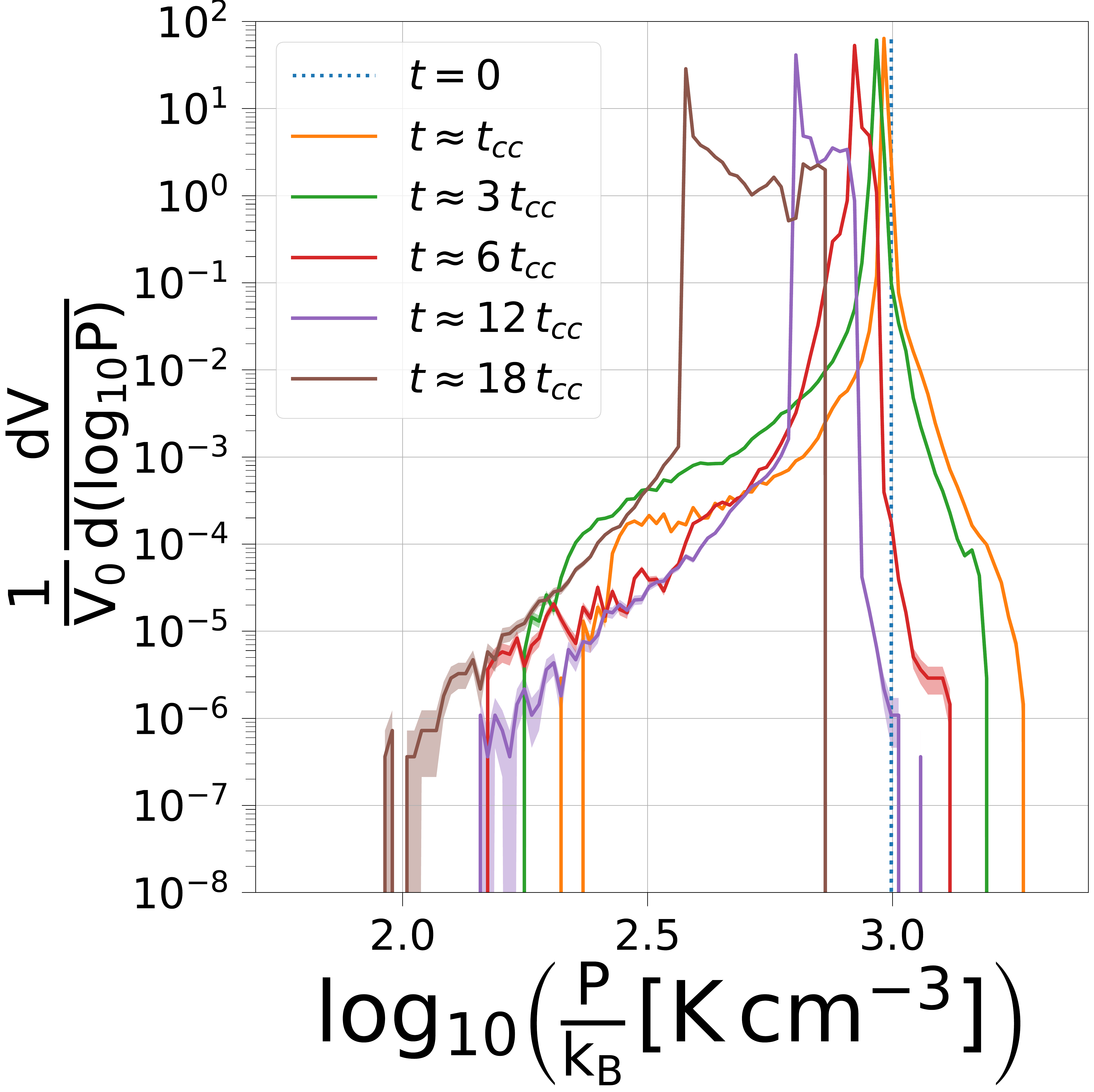}
\end{subfigure}
\caption{Some useful PDFs at different times for our fiducial simulation at the resolution of $R_{\rm cl}/d_{\rm cell} = 8$ (see Table~\ref{tab:initial_radius}). The counting uncertainty in each bin of the PDFs is assumed to follow a Poisson statistic and hence a shaded spread $\propto 1/\sqrt{\rm bin\ count - 1}$ is marked for every bin. \textbf{Left: }The mass-weighted PDF (mass-filling fraction) of temperature shows the cold and the hot phases. Because of the large box-sizes in our simulations, the mass fraction of the hot phase is much higher than the cold phase. With time, turbulent mixing and cooling create intermediate temperature phases. In absence of cooling, the cold peak mixes into the hot one in a few $t_{\rm cc}$. Background cooling results in the slight shift of the hot phase peak to lower temperature values at late times. \textbf{Middle: }Time evolution of the volume-weighted PDF (volume-filling fraction) of temperature. \textbf{Right: }The volume-weighted PDF of pressure shows only one peak at all times. This shows that mixing and cooling in the boundary layer occur in approximately isobaric conditions. The low pressure tail corresponds to the cool/mixed gas formed in the turbulent cooling wake. The high pressure tail at early times is produced mainly by the gas crossing the bow shock and at late times by the transient shocks forming around cooling blobs (see the right panels of Figure~\ref{fig:slices}). Once again, note that background cooling drives the pressure peak to smaller values.
}
\label{fig:1d-pdfs}
\end{figure*}
\begin{figure}
  \centering 
  \includegraphics[width=\linewidth]{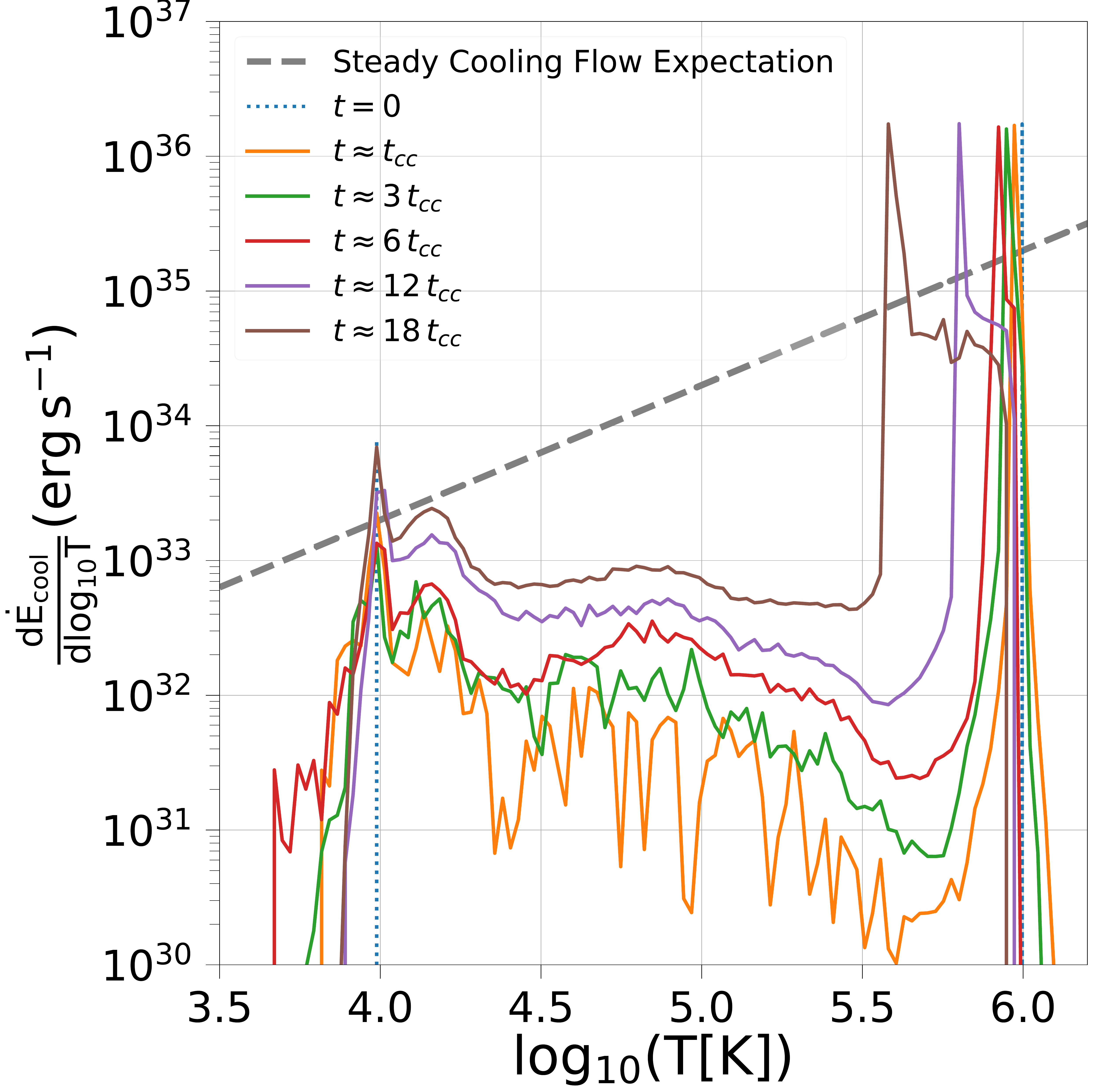}
\caption{The differential cooling rate at different temperatures for our fiducial simulation at a resolution of $R_{\rm cl}/d_{\rm cell} = 8$ (see Table~\ref{tab:initial_radius}) . The hot gas, which fills most of the simulation volume, naturally causes a large peak at the high temperature. This peak slightly shifts to the left with time due to the cooling of the background hot gas. Cooling losses at intermediate temperatures are significant, and the features in the differential cooling losses reflect the interplay between turbulent mixing and cooling moving the gas across temperatures. The differential emission measure is qualitatively different from a simple steady cooling flow (as shown by the dashed line corresponding to Eq. \ref{eq:CF}; see section~\ref{sec:cooling_flow} for details). }
\label{fig:DEM}
\end{figure}

Probability distribution function (PDFs) are commonly used to describe the statistical properties of turbulent flows (e.g., \citealt{pope2000}). Figure~\ref{fig:1d-pdfs} shows some useful PDFs at different times for the fiducial cloud-crushing simulation with $R_{\rm cl}/d_{\rm cell}=8$ (see Table~\ref{tab:initial_radius}). The left panel  shows the mass-weighted temperature PDF (mass filling fraction), the middle panel shows the volume-weighted PDF (volume filling fraction), and the right panel shows the volume PDF of pressure. Figure \ref{fig:DEM} shows the differential cooling rate at different temperatures in the simulation domain.
The temperature PDFs appear as two sharp peaks at $t=0$. The amplitude of the low-temperature peak in the temperature PDFs (marking the cold gas) falls and broadens during the early times because of turbulent mixing. This cold peak undergoes a significant rise at later times due to the cooling of the mixed gas and the subsequent cold mass growth. The volume and mass of the gas at intermediate temperatures also rise with time due to turbulent mixing and cooling. 

Notice that the hot peak moves to the lower temperatures (accompanied by a movement to low pressure as seen in the pressure PDF) as the runtime approaches the cooling time of the diffuse hot wind. After this time we expect the whole computational box to cool to the cooling floor. While the cooling time of the background hot gas in the central CGM can be much shorter than the Hubble time and it is important to study its thermodynamics (e.g., \citealt{sharma2018}), in this paper  we only focus on the cooling of the turbulent boundary layers. 

The pressure PDF, unlike temperature, has a single peak, indicating that the conditions remain nearly isobaric. Although localized regions at intermediate temperatures can cool faster than the sound-crossing time, for most of the volume and time cooling is subsonic. While \citet{Fielding_2020} have argued that the low pressure in the cold layer is an artifact of insufficient resolution, our pressure PDF shows a similar low-pressure tail even at higher resolutions (see the right panel of Fig. \ref{fig:pdf-high-res}). A resolved low-pressure tail at late times is also clearly visible in the bottom-right panel of Figure \ref{fig:slices} corresponding to our highest resolution fiducial simulation.

Figure~\ref{fig:DEM} shows the radiative cooling losses (assuming optically thin conditions) as a function of the gas temperature. We obtain $d\dot{E}_{\rm cool}/d\log_{10}T$ by calculating the cooling losses $\dot{E} = \int_V n^2\Lambda(T)dV$ as a function of temperature distributed across uniform bins in $\log_{10}T$. The radiative cooling PDF also has two peaks at all times, corresponding to the hot and the cold phases. Although the mass and volume in the intermediate temperatures is low, the cooling rate at these temperatures is significant because the cooling function peaks at these temperatures ($\sim 10^5$ K).

\subsubsection{Comparison with a simple cooling flow}
\label{sec:cooling_flow}
\begin{figure}
  \centering 
  \includegraphics[width=\linewidth]{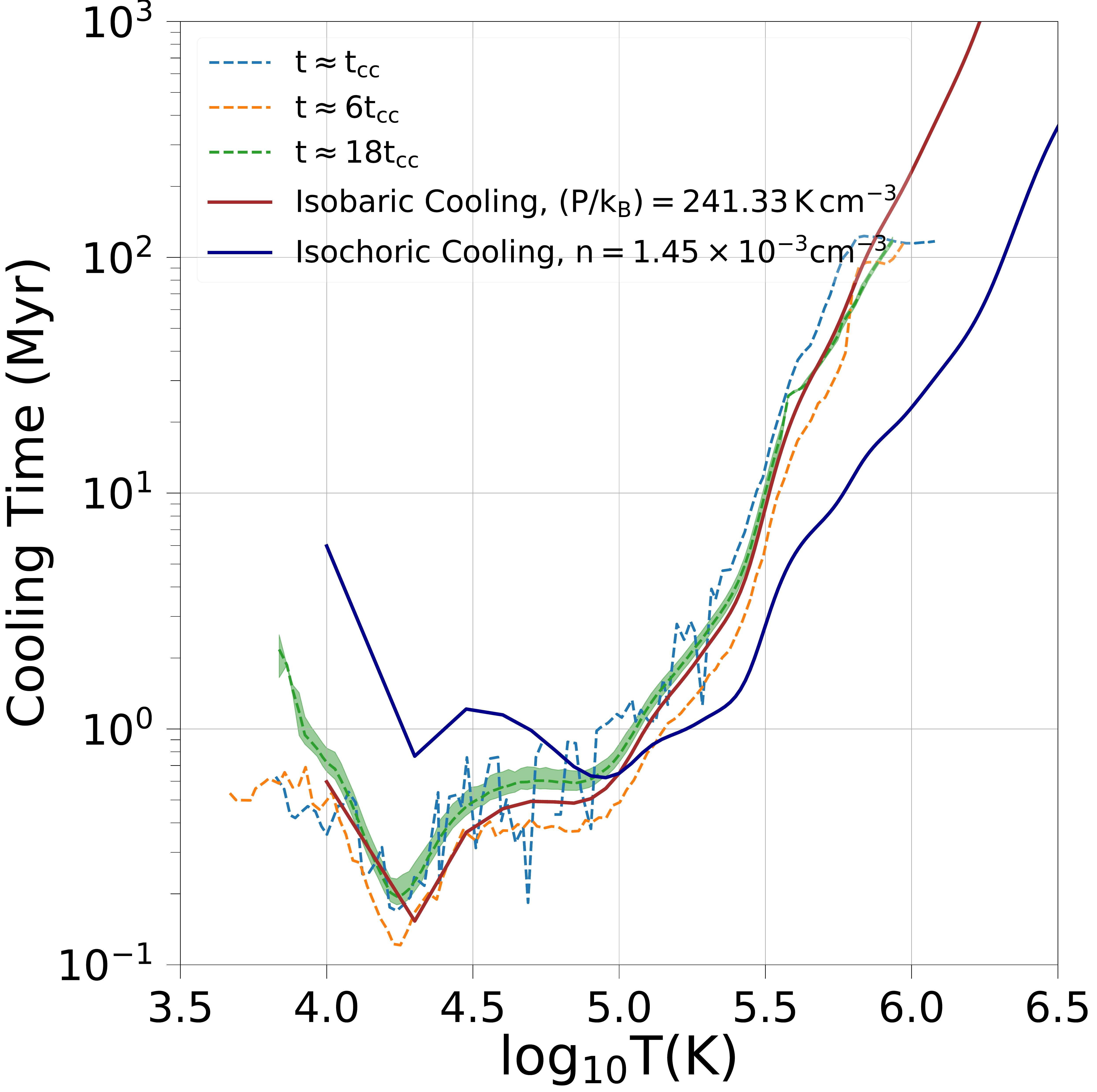}
\caption{The median cooling times at different temperatures for the fiducial simulation ($R_{\rm cl}/d_{\rm cell}=8$; see Table~\ref{tab:initial_radius}) at different times are represented by the dashed lines. For $t\approx 18 t_{\rm cc}$ we also show the $1-\sigma$ spread around the median value. The solid lines show the analytic isobaric (Eq.~\ref{eq:tcool_IB}) and isochoric (Eq.~\ref{eq:tcool_IC}) cooling times as a function of temperature, with the pressure adjusted to match the simulation results (the density for the isochoric cooling time is adjusted to cross the isobaric cooling time at $10^5$ K). A close match of the simulation results with the isobaric cooling time reflects the near isobaric conditions in the turbulent cooling wake, although cooling time from the simulation indicates a lower pressure toward lower temperatures. 
}
\label{fig:cooling_isob}
\end{figure}
Figure~\ref{fig:cooling_isob} shows the analytic isobaric and isochoric cooling times as a function of temperature,
\begin{equation}
\label{eq:tcool_IB}
    t_{\rm cool, IB} = \frac{5}{2 p} \frac{k_B^2 T^2}{\Lambda(T)}
\end{equation}
and
\begin{equation}
\label{eq:tcool_IC}
    t_{\rm cool, IC} = \frac{3}{2} \frac{k_BT}{n\Lambda(T)}.
\end{equation}
For the figure, the pressure in the isobaric cooling time (Eq. \ref{eq:tcool_IB}) is chosen to roughly match the pressure in the mixing layer at $18 t_{\rm cc}$. The density for the isochoric cooling time (solid blue line) is adjusted to cross the isobaric cooling time (solid red line) at $\sqrt{T_{\rm cl}T_{\rm hot}}=10^5$ K. We also plot the median cooling time across the computational domain as a function of temperature at different times (dashed lines). We shade the $1-\sigma$ spread of the cooling time for $t\approx18 t_{\rm cc}$. 

We find that the cooling time from the simulation matches very well with the isobaric cooling time using a pressure lower than the initial pressure. This reflects the smaller pressure and roughly isobaric conditions in the cooling turbulent boundary layer. Notice closely in Figure \ref{fig:cooling_isob} that the analytic isobaric cooling time lies above the cooling time from the simulation close to $10^6$ K and progressively below it as one goes to lower temperatures, indicating that the lower temperature gas has increasingly lower pressure till $\approx 10^{4.2}$ K. The various bumps and wiggles in the cooling time reflect the similar features in the cooling curve. Note that the cooling time versus temperature is not as smooth at early times but the variations are smoothed out with time.

While Figure~\ref{fig:cooling_isob} clearly indicates isobaric conditions in the cooling boundary layer, the energy radiated by gas at different temperatures in Figure~\ref{fig:DEM} shows that this is not a simple isobaric cooling flow (\citealt{fabian1994}). For a single-phase (not clumpy) steady cooling flow, the energy loss rate in a given temperature bin $\Delta \dot{E}_{\rm cool} = (d\dot{E}_{\rm cool}/d\log_{10} T) \Delta \log_{10} T \approx \Delta E/t_{\rm cool}$ has one-to-one correspondence with the constant mass cooling rate in the temperature bin $\Delta \dot{M}_{\rm cool} \approx \Delta M/t_{\rm cool}$; $\Delta \dot{E}_{\rm cool}/\Delta \dot{M}_{\rm cool} \approx \Delta E/\Delta M \approx 5k_BT/(2\mu m_p)$ (assuming isobaric conditions; $\mu m_p$ is the mean particle mass, and $\Delta E$ and $\Delta M$ are the enthalpy and mass in the temperature bin). This implies that for a steady cooling flow (i.e., a constant $d\dot{M}_{\rm cool}/d\log_{10}T$ across temperatures),
\begin{equation} 
\label{eq:CF}
\frac{d\dot{E}_{\rm cool}}{d \log_{10} T} \approx 2\times 10^{35} {\rm~erg~ s}^{-1} \left( \frac{T}{10^6{\rm~K}} \right) \left( \frac{d\dot{M}_{\rm cool}/d\log_{10} T}{10^{-5} M_\odot {\rm yr}^{-1}} \right).
\end{equation}

The dashed line in Figure~\ref{fig:DEM} shows the above scaling for $d\dot{M}_{\rm cool}/d\log_{10}T \approx 10^{-5} M_\odot~{\rm yr}^{-1}$ measured from the average growth rate of dense gas ($\rho <3 \rho_{\rm cl})$ in the almost steady growth phase after $\approx 6 t_{\rm cc}$ (see the top panel of Fig. \ref{fig:glob-prop-GO}). Clearly the differential emission measure obtained from the simulation is different (even qualitatively) from the steady cooling flow model. Turbulence in the boundary layer moves mass across temperature bins in both directions, breaking the close correspondence between $\Delta \dot{M}_{\rm cool}$ and $\Delta \dot{E}_{\rm cool}$ of a cooling flow.

The emissivity at $10^{5.5}$ K is two orders of magnitude smaller than the steady cooling flow model with the steady mass growth rate of dense gas. Moreover, the emissivity is flatter, with excess radiative losses at low and intermediate temperatures. This behavior is rather similar to the differential emission measure seen in the idealized cluster simulations with multiphase gas (Figure 3 in \citealt{sharma2012}), both for a multiphase cooling flow and for a suppressed cooling flow in presence of feedback heating. This suggests that a single-phase cooling flow is almost never realized (even in a steady state) in multiphase flows with cooling (see also Fig. 14 in \citealt{mandelker2020}). This fundamental property of turbulent, multiphase boundary layers may explain the diverse range of ions seen in the CGM observations. 

\subsubsection{Joint probability distribution functions}\label{sec:2d-pdfs}
\begin{figure}
  \includegraphics[width=\columnwidth]{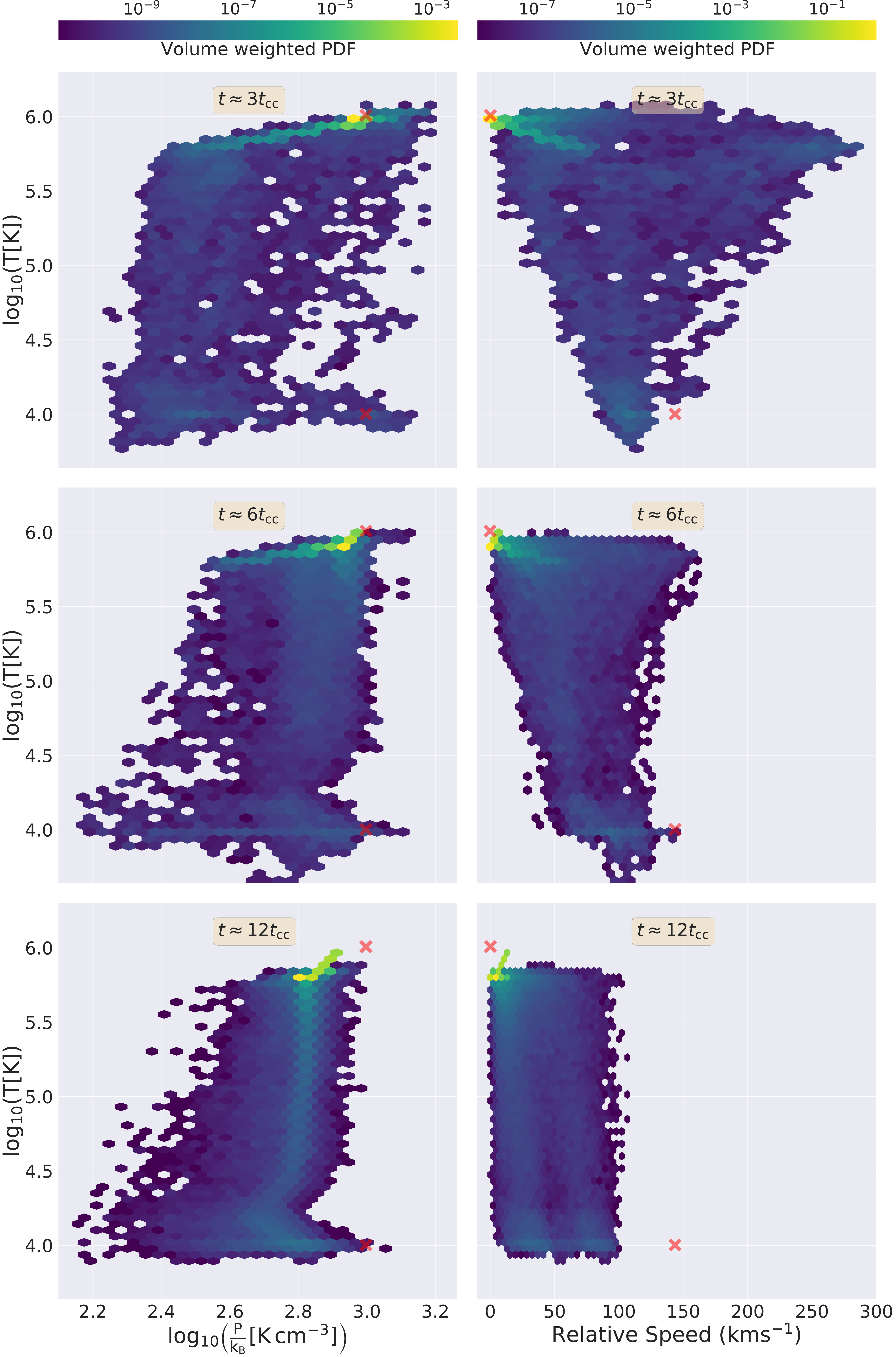}
\caption{The joint probability distributions (volume-weighted) of pressure-temperature (left panels) and relative speed-temperature (relative to the hot wind material; right panels) at different times for the fiducial run with $R_{\rm cl}/d_{\rm cell}=8$. From the top row to the bottom, the plots are made at \{3,6,12\} $t_{\rm cc}$. The pressure does not exhibit a lot of variation, especially at late times, indicating an approximate isobaric evolution. The hot wind starts to entrain the cold gas due to turbulent momentum transport by about a drag time $\sim 10 t_{\rm cc}$, consistent with the right panel of Figure~\ref{fig:glob-prop-GO}. The red crosses indicate the position of the peaks corresponding to the initial ($t=0$) probability distribution.
}
\label{fig:2d-pdf}
\end{figure}

Figure~\ref{fig:2d-pdf} shows some two dimensional (volume-weighted) PDFs of pressure-temperature and relative speed-temperature (relative speed is calculated with with respect to the instantaneous rest frame of the wind material) for all the grid cells in our fiducial simulation ($R_{\rm cl}/d_{\rm cell}=8$; see Table~\ref{tab:initial_radius}) at different times. The red crosses indicate the locations of peaks in the initial PDFs. At $t=0$ the cold cloud and the hot wind have the same pressure, temperatures differing by $\chi=100$, and speeds differing by $v_{\rm wind}\approx$150 km s$^{-1}$.

In absence of radiative cooling (not shown) the low temperature peak completely merges into the hot peak by a few $t_{\rm cc}$ (see the dotted lines in the top panel of Figure~\ref{fig:glob-prop-GO}). However, with cooling and for $R_{\rm cl}>R_{\rm GO}$, the cold gas is replenished by the cooling of the mixed phase and hence the peak at $T_{\rm cl}$ is seen at all times. The spread in pressure values of all the fluid elements as illustrated in the joint PDFs the joint PDFs (left panels) is larger at early time because of a strong bow shock, rarefaction, and fast cooling of the mixed gas (see the right panels of Figure~\ref{fig:slices}). At late times, the pressure is more uniform, with a slight dip close to the cloud temperature, indicating a pressure-driven turbulent cooling flow on to the dense tail.

The high velocity peak in the relative speed-temperature PDF slowly moves to lower velocities and fully crosses 100 km s$^{-1}$ by 12 $t_{\rm cc}$, which is of order the drag time. Thus, in agreement with the bottom panel of Figure~\ref{fig:glob-prop-GO}, the cold/dense gas starts to become comoving with the hot wind after roughly a drag time. Once the hot and cold phases do not have a relative speed, the turbulent cooling boundary layer is expected to decay as shear driven turbulence gradually weakens.

\subsection{Criterion for cloud growth}
\label{sec:discrepancy}

One of the aims of this paper is to reconcile the large discrepancy in the Gronke-Oh and Li criteria, expressed most transparently in terms of a threshold cloud radius (Eq.~\ref{eq:GO_radius} and Eq.~\ref{eq:Li_radius}; Figure~\ref{fig:GOLidep}) for the growth of cold/dense gas in the cloud-crushing problem with cooling. For this, we have carried out simulations with cloud radii in between the Gronke-Oh and Li radii. We have not extensively tested the exact value of the threshold cloud radius for growth because 3D simulations with long boxes and sufficiently high resolution are expensive, and cold gas takes longer to grow as we move closer to the threshold (see the top panel of Figure~\ref{fig:glob-prop-GO}).

Figure~\ref{fig:Overlay} clearly shows that our simulations with cloud radii between the Gronke-Oh and Li radii show growth of cold gas, and thus they are consistent with \citetalias{gronke2018}, \citetalias{gronke2020} but not with \citetalias{Li2020}. This figure is based on a similar figure in \citetalias{Li2020} (their Figure 3) but also populated by data points from our simulations with radii smaller than the Li radius (Eq.~\ref{eq:Li_radius}). The solid lines show the Gronke-Oh criterion ($n_{\rm cl}=0.1$ cm$^{-3}$) for different Mach numbers in the same parameter space as used by \citetalias{Li2020}, which differ greatly from the Li criterion. We may justify the different analytic growth criteria because they are based on different physical arguments. However, the results of different numerical simulations of radiative cloud-crushing must agree as they essentially simulate the same setup with very similar parameters. Now we discuss possible causes for the large discrepancy between the numerical results of the different groups. 

We note that while \citetalias{gronke2018,gronke2020} run radiative hydrodynamic simulations like us, simulations of \citetalias{Li2020} include magnetic fields, thermal conduction/viscosity, as well as self-shielding and self gravity, and their cloud survival criterion is based on these simulations. So  one may not directly compare the Gronke-Oh and Li criterion. However, \citet{Sparre2020} carry out MHD simulations of cloud-crushing without conduction and other effects included by \citetalias{Li2020} but find a threshold consistent with \citetalias{Li2020} (except for a strong Mach number dependence), hinting that the Li threshold is governed mainly by the interplay of hydrodynamic mixing and cooling. Further simulations are necessary to isolate the quantitative effects of the additional physics of magnetic fields and thermal conduction for a range of parameters.

A central question here is: {\it how long should the cloud-crushing simulations be run to assess the cooling of the turbulent mixing layer?} The answer is that we should run for at least an appreciable fixed fraction (say 0.5; this corresponds to $18t_{\rm cc}$ for our fiducial parameters) of the cooling time of the hot phase (after this the background medium cools, which is not the regime of interest in the cloud-crushing problem). An unambiguous way to assess whether dense gas grows is to run the cloud-crushing simulations as long as possible, which in the present case means until the background medium starts to cool.\footnote{While the hot gas cooling time is irrelevant for the physics of the radiative cloud-crushing problem as the hot is brought to intermediate temperatures by turbulent mixing rather than radiative cooling, we should run the simulations as long as possible to check dense mass growth at late times.} Since the cooling time of the hot phase $t_{\rm cool, hot} \propto T_{\rm cl} \chi^2/[n_{\rm cl} \Lambda(\chi T_{\rm cl})]$ increases faster with $\chi$ (e.g., see  Figure~\ref{fig:cooling_isob}) than the cloud-crushing time $t_{\rm cc} \propto \chi^{1/2} R_{\rm cl}/v_{\rm wind} \propto \chi^{1/2}/\mathcal{M}$, the simulations with larger density contrast ($\chi$) and Mach number ($\mathcal{M}$) must be run for more cloud-crushing times ($t_{\rm cc}$) to see if dense gas grows due to cooling of the mixing layer. A longer run-time for higher $\chi$ and $\mathcal{M}$ also demands a larger box-size because the mixed gas has more time to leave the computational domain through the boundaries. 

\citetalias{Li2020} declare a cloud destroyed when its mass falls below 10\% of its initial mass but the dense gas may still grow  even after such a large dip. Ideally, the simulation time should be as long as possible until the catastrophic cooling of the background wind. Very recently, \citet{Sparre2020} have performed numerical simulations of the cloud-crushing problem with cooling (their resolution $R_{\rm cl}/d_{\rm cell}=7$ for $\chi=100$, similar to ours) and mapped out the parameter regime for the growth of cold gas by monitoring the cold gas mass till $12.5 t_{\rm cc}$ for most of their runs, which is somewhat arbitrary. Figure A2 of \citet{Sparre2020} shows that their simulations with $R_{\rm cl} = 15\,\rm pc$ and $R_{\rm cl} = 47\,\rm pc$ (for $\chi=100$ and $\mathcal{M}=1.5$) show growth of cold gas after $12 t_{\rm cc}$ but before catastrophic cooling of the background. Therefore these simulations are in the growth regime according to our robust definition of cloud growth. Thus the simulations of \citet{Sparre2020} (perhaps even \citetalias{Li2020}) show growth of dense gas for clouds smaller than the Li radius and there is broad agreement among simulation results. More work is required for a more precise comparison.

Now we move on to other factors that can affect the outcome of radiative cloud-crushing simulations such as the box-size and resolution. The typical simulation domain chosen by \citetalias{Li2020} is smaller ($20 \times 10\times 10 R_{\rm cl}^3$) than our fiducial box-size ($400\times 30 \times 30 R_{\rm cl}^3$). For a small box, the mixed/dense gas can leak out of the simulation domain and can suppress the formation of cool gas seeds, which leave the simulation box rather than seed the growth of cold gas in the long cloud tail. Such a cold cloud would grow in larger boxes. We explore the effects of the box-size and the boundary conditions in Appendix~\ref{app:box-size}. We indeed verify that a smaller simulation box gives a larger threshold radius for the growth of cold gas (the bottom panel of Figure~\ref{fig:dme_boxbcsizecrir}) but the effect just on its own is not big enough to explain the large difference between the Gronke-Oh and Li radii. For example, a small box-size of $(20,10,10) R_{\rm cl}$ for $R_{\rm cl}/d_{\rm cell}=8$ and fiducial parameters ($\chi=100$, $\mathcal{M}=1.0$) shows cloud growth only for the cloud-size interpolation parameter $\zeta >  0.6$,  whereas the large box simulation for the same physical parameters shows cloud growth even at $\zeta=0.5$  (corresponding to the geometric mean of Gronke-Oh and Li radii; see Eq.~\ref{eq:zeta}).

The top panel of Figure~\ref{fig:glob-prop-GO} shows that in some of our simulations, especially closer to the threshold radius for growth, there is a significant dip in the cold mass at early times before the cloud starts to grow. Moreover, in the left panel of Figure~\ref{fig:dme_res} we find that the time till the dip in cold gas mass is longer for a higher resolution with a larger dip in dense mass (Appendix~\ref{app:res} investigates the effect of resolution). Note that \citetalias{Li2020} use a mass resolution of $\sim 10^{-6} M_{\rm cl}$, which corresponds to $R_{\rm cl}/d_{\rm cell} \sim 64$ in the dense phase, $\sim 40 (\chi/100)^{-1/6}$ in the intermediate phase and $\sim 14 (\chi/100)^{-1/3}$ in the hot phase, much higher than our (and that of \citealt{Sparre2020}) typical resolution $R_{\rm cl}/d_{\rm cell}=8$. The red line in the left panel of Figure~\ref{fig:dme_res}, which corresponds to $R_{\rm cl}/d_{\rm cell}=64$ and a box-size of $(30,15,15)R_{\rm cl}$, shows a dip in the dense mass fraction below 0.1. So the cloud in this run, according to the definition used by \citetalias{Li2020}, will be labelled as destroyed. Since the box-size of \citetalias{Li2020} is even smaller ($[20,10,10]R_{\rm cl}$), the drop in dense mass fraction will be even more. Thus, it possible that \citetalias{Li2020} do not run their simulations long enough to observe the cloud growth after the long dip in the cold gas mass for a large $R_{\rm cl}/d_{\rm cell}$. So, a combination small box-size, their definition of cloud destruction, high resolution and small run time may lead them to misidentify some of the growing clouds as destroyed.

Apart from these factors, the mismatch in the simulation results may be due to the difference between the numerical algorithms. \citetalias{Li2020} use the mesh-free smooth particle hydrodynamic code GIZMO (\citealt{hopkins2015}) while \citetalias{gronke2018}, \citetalias{gronke2020} use a finite volume Godunov code  ATHENA (similar to PLUTO; \citealt{stone2008}). Generally fixed-grid codes overestimate numerical mixing which may lead to the differences in the numerical results among various groups. Thus a quantitative resolution of \citetalias{gronke2018} and \citetalias{Li2020} results and mapping of the precise growth threshold require further work and a closer comparison.
\section{Caveats \& future directions}
\label{sec:caveats}

While the discovery of a criterion for the growth of cold gas in cooling turbulent boundary layers is an important achievement (\citetalias{gronke2018}), there are several conundrums still remaining to be solved when applying this idea to the observations of the CGM and galactic outflows. Even before that, the exact threshold needs to be mapped out with high resolution 3-D simulations of large enough boxes.

The Gronke-Oh criterion is based on the implicit assumption that the cooling function peaks at the intermediate temperatures. Moreover, estimating the cooling time of the mixing layer as $\Lambda(T_{\rm hot}^{1/2}T_{\rm cl}^{1/2})$ is not rigorously justified. This important estimate will clearly depend on the shape of the cooling function. The effective cooling function can be quite different, for example, in AGN outflows irradiated by intense radiation (\citealt{dyda2017}).

In several multiphase media, the cooling time of the hot diffuse phase is shorter than the timescales of interest; e.g., the inner CGM (e.g., \citealt{cavagnolo2009}) and the central parts of galactic outflows (e.g., \citealt{thompson2016}). In this case, most of the cold gas can be produced by the cooling of the hot gas with large density perturbations (\citealt{choudhury2019}). The cloud-crushing studies assume a pre-existing dense cloud, the origin of which is non-trivial. The cold gas can condense out of the hot CGM if the ratio of the cooling time to the free-fall time is less than a threshold $\lesssim 20$ (\citealt{voit2015,lakhchaura2018}). Cold clouds can be pushed up and entrained by superbubbles blowing out of the gas disk (\citealt{vijayan2018,schneider2020}) but they seem to get destroyed by $\sim 10$ kpc, implying that the cold gas growth in the turbulent boundary layers depend crucially on the background conditions and is not yet fully understood. Ram pressure stripped galactic wakes (e.g., \citealt{ebeling2014}) are another way to seed cool gas that can cool and seed growing cold clouds (\citealt{tonnesen2010,yun2019,nelson2020}).

In the standard cloud-crushing problem, the cold gas eventually becomes comoving and the turbulent cooling boundary layer does not exist any more. In order to quantitatively apply cloud-crushing results to CGM clouds, the background gravity must be taken into account. The question of the survival of the IGM filaments penetrating into the CGM of halos of different masses (\citealt{dekel2009}) is also intimately connected to the interplay of turbulent boundary layers and cooling (\citealt{mandelker2020,Fielding_2020}).

Apart from the various issues discussed above, there are the usual concerns about the convergence of numerical results in absence of explicit dissipation (e.g., \citealt{koyama2004}; but see \citealt{Tan2020}). Important physical effects such as thermal conduction, magnetic fields, background turbulence (e.g., \citealt{mohapatra2019,li_y2020}), and cosmic rays (e.g., \citealt{butsky2020}) can definitely affect the small scale structure of the multiphase plasma. This rich problem clearly deserves to be studied systematically, carefully benchmarking the effects of these various important effects. Only after this, can we confidently apply the cloud-crushing results to the observed multiphase plasmas.

\subsection{Hot wind cooling time \& cloud growth}

\begin{figure}
\centering
	\includegraphics[width=0.86\columnwidth]{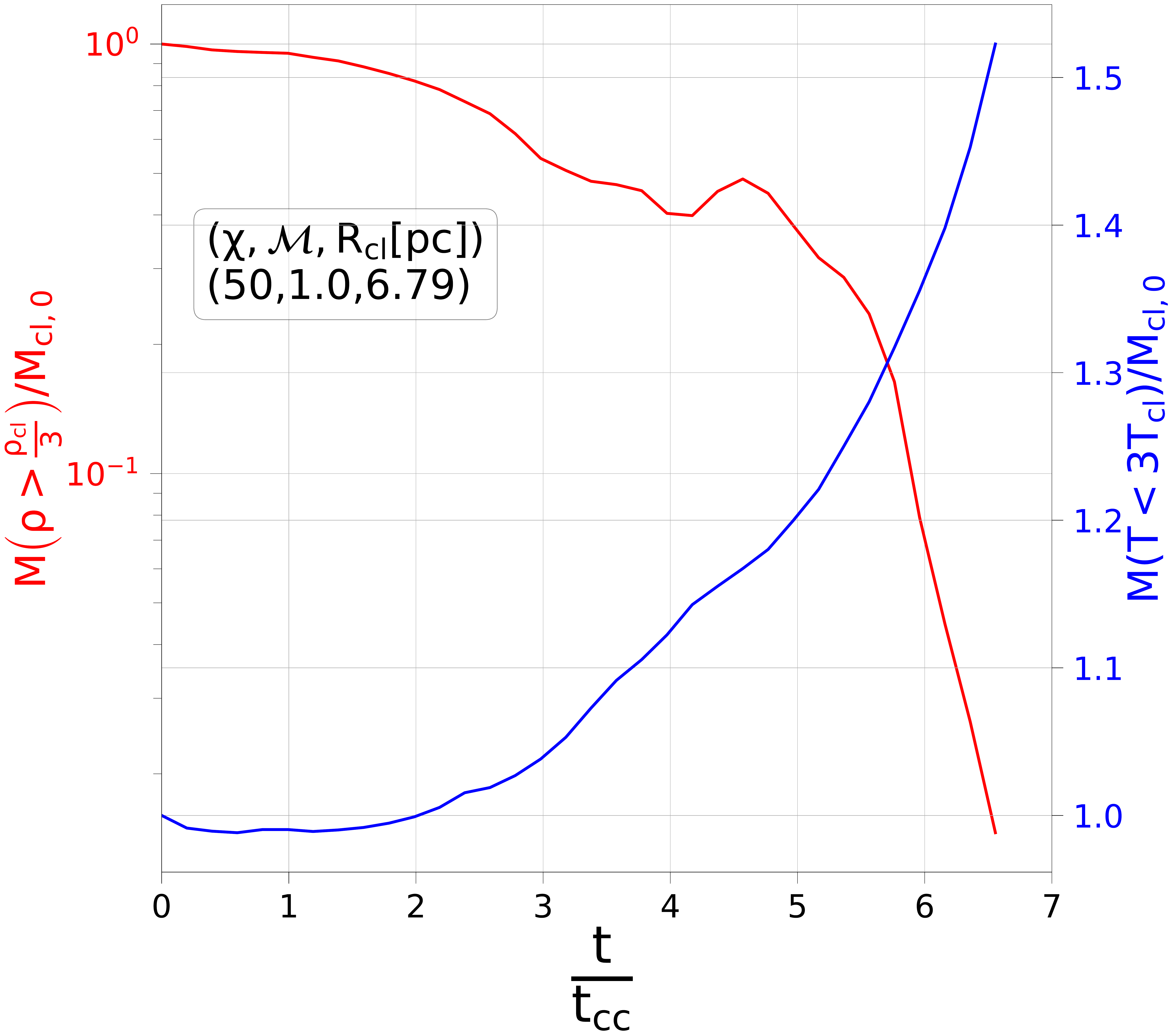}
	\includegraphics[width=0.995\columnwidth]{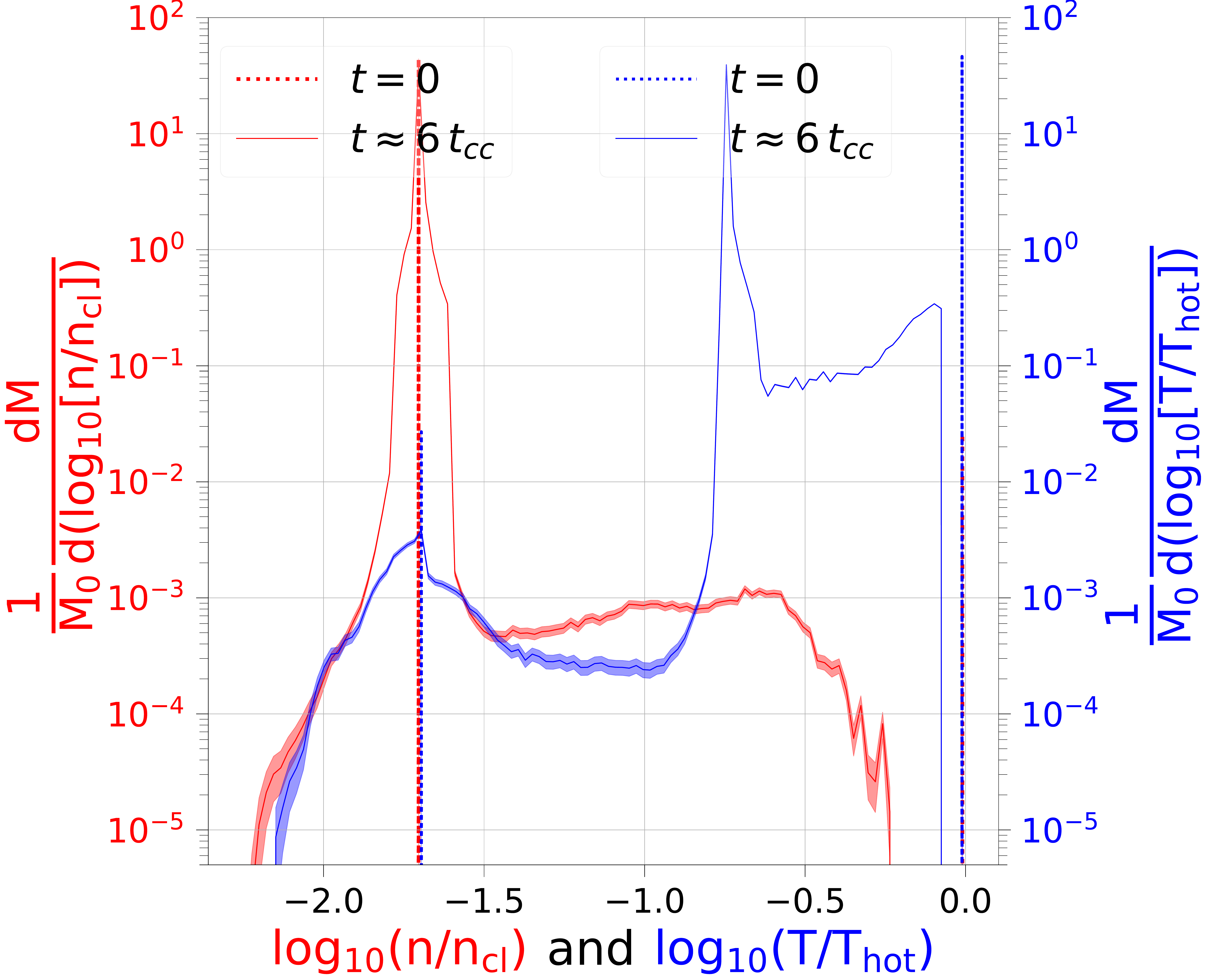}
    \caption{The top panel shows the dense ($\rho>\rho_{\rm cl}/3$; in red) and cold ($T<3T_{\rm cl}$; in blue) mass fraction as a function of time and the bottom panel shows PDFs of  density (red) and temperature (blue) at $t=6t_{\rm cc}$ (initial PDFs are shown by dotted lines) for the run with a density contrast $\chi=50$ (see Table~\ref{tab:initial_radius}). The dense ($\rho>\rho_{\rm cl}/3$) mass evolution in the top panel incorrectly suggests that the cloud much larger than the Gronke-Oh radius ($\approx 1$ pc; Eq.~\ref{eq:GO_radius}) is destroyed with time. On the other hand, the cold ($T<3T_{\rm cl}$) mass grows with time. This contrasting behavior is because of cooling of the hot wind and the expansion of the dense cloud in presence of the reduced ambient pressure. The mass PDF of density (normalized to $n_{\rm cl}$) at $6 t_{\rm cc}$ (in red) confirms that the peak cold density shifts to lower than $n_{\rm cl}/3$. The hot gas temperature peak (in blue) moves to $<T_{\rm hot}/3$, indicating a decrease in the confining pressure. The hot gas density and the cold gas temperature at $6 t_{\rm cc}$ still peak at their initial values. Once again, a spread $\propto$ $\rm 1/\sqrt{bin\ count-1}$ is shown for the PDFs.}
    \label{fig:chi_50}
\end{figure}

As an example of caution that we need to exercise in order to apply the cloud-crushing results, we consider the cloud-crushing problem with cooling for a density contrast of $\chi=50$. In this case, the background cooling time is not much longer than the cooling time of the mixed gas (see Figure~\ref{fig:cooling_isob}). Figure~\ref{fig:chi_50} shows the dense ($\rho>\rho_{\rm cl}/3$) and cold ($T<3T_{\rm cl}$) mass evolution for a cloud at $\chi$=50, $\mathcal{M}=1.0$, and a radius of $R_{\rm cl}\approx 8 R_{\rm GO}$ (see Eq.~\ref{eq:GO_radius}). From the left panel, we can clearly see that the dense mass is `destroyed' after $\approx 6 t_{\rm cc}$, with the evolution similar to the cases where clouds are destroyed due to inefficient cooling in the mixing layer. However, the evolution of the {\it cold} ($<3 T_{\rm cl}$) gas mass shows growth. How is it that the cold mass grows but the dense mass decreases? The reason is the cooling of the background hot wind, which decreases the ambient temperature and pressure. In a low ambient pressure, the cold gas at $\sim 10^4$ K expands isothermally and reduces its density to $\rho<\rho_{\rm cl}/3$ but the gas at $T<3T_{\rm cl}$ keeps building up with time. 

The importance of the $\chi/\mathcal{M}$ dependence of the hot wind cooling time and the appropriate time-window to analyze cloud growth in the boundary layer, presented as the key cause of discrepancy between the Gronke-Oh and Li criteria for cloud growth in section~\ref{sec:discrepancy}, is reinforced by this example. 

\section{Summary}
\label{sec:summary}
In this paper, we systematically analyze the survival of cold gas in the cloud-crushing problem in presence of optically thin radiative cooling, particularly for the parameters relevant to the CGM. Our conclusions are summarized as follows:
\begin{enumerate}
	\item We analyze the large discrepancy (see Figure~\ref{fig:GOLidep}) between the cloud growth criteria predicted by \citetalias{gronke2018}, \citetalias{gronke2020} and \citetalias{Li2020}. In our numerical simulations, clouds smaller than the Li radius and larger than the Gronke-Oh radius show the growth of cold gas (see Figure~\ref{fig:Overlay}). The main reason that \citetalias{Li2020} and \citet{Sparre2020} see cloud destruction for smaller clouds is their somewhat arbitrary definition of cloud destruction. We propose a robust definition of cloud growth for the radiative cloud-crushing problem. A smaller computational domain also suppresses cloud growth.
	Simulations with larger density contrast and Mach numbers should be run for more cloud-crushing times because the cooling time of the hot wind is longer. Thus, our results stress the importance of the cooling time of the {\it mixed} gas rather than the hot gas for the growth of dense mass. However, more work is needed to map out the exact cloud growth threshold with large boxes and sufficient resolution. It is also unclear whether the most appropriate temperature of the mixing layer to evaluate its cooling time is $\sqrt{T_{\rm hot}T_{\rm cl}}$ or something else depending on the exact cooling curve.
    \item The volume and mass probability distribution functions (PDFs) of temperature, pressure, emissivity, and velocity (Figs. ~\ref{fig:1d-pdfs}, \ref{fig:DEM} and \ref{fig:2d-pdf}) are useful diagnostics to understand the structure and dynamics of the cooling turbulent boundary layer. The density and temperature PDFs are bimodal but pressure has a relatively narrow distribution with a low-pressure tail, reflecting a roughly isobaric turbulent boundary layer which grows due to cooling and entrainment of the mixed gas.
    \item The cooling time-temperature distribution in the boundary layer very closely tracks the isobaric cooling time at the ambient pressure (Figure~\ref{fig:cooling_isob}). However, the temperature distribution of the emissivity is different from a simple homogeneous cooling flow (see section~\ref{sec:cooling_flow}), implying that the movement of mixed gas across intermediate temperatures is not simply because of radiative cooling. A broad and a flat emissivity distribution with temperature appears to be generic for radiative multiphase flows. This may  explain the diverse ions seen in the CGM observations.
    \item One should be careful about the validity of the cloud-crushing results. E.g., thermodynamics of the diffuse hot background becomes very crucial on timescales longer than the cooling time of the ambient hot medium. A very important question that the cloud-crushing setup bypasses is that of the origin of large enough cold clouds in the first place (see section~\ref{sec:caveats}).
	\item Although the global properties of the multiphase gas in our simulations agree qualitatively, their detailed morphology depends on resolution and does not show convergence up to the highest resolutions we simulate ($R_{\rm cl}/d_{\rm cell}=64$). 
	More simulations are needed to precisely map out the threshold radius for the growth of dense clouds as a function of the key parameters (e.g., $\chi$ and $\mathcal{M}$).
\end{enumerate}
 
\section*{Acknowledgements}
We acknowledge Ryan Farber, Max Gronke, Zihui Li, Nir Mandelkar, Peng Oh, Martin Sparre, Enrique Vazquez Semadeni, and the anonymous reviewer for useful comments and discussions. We acknowledge the support from the Supercomputer Education and Research Centre (SERC) at the Indian Institute of Science (IISc), Bangalore where most of the production runs of our simulations were carried out. VK and AD would like to express their gratitude to the super-responsive and helpful people at the \href{https://discourse.paraview.org/}{Paraview forum} and the \href{https://developer.nvidia.com/nvidia-index}{Nvidia IndeX team} for all the technical support needed for some of the visualizations. Research of VK is supported by the Prime Minister's Research Fellowship (PMRF)  from the MHRD, Govt. of India. AD acknowledges the support from the Max Planck Institute for Astrophysics, Garching for accepting him as a visiting student and providing access to their Freya supercomputing facility where initial part of the code development and prototyping were carried out. PS acknowledges a Swarnajayanti Fellowship from the Department of Science and Technology (DST/SJF/PSA-03/2016-17).

\section{Data Availability}
All the relevant data associated with this article will be shared on reasonable request to the authors. Some of the relevant plotting and visualization codes will soon be hosted on \href{https://github.com/IISc-Computational-Astrophysics/Cloud-Crushing-2020}{Github} for public access.

\bibliographystyle{mnras}
\bibliography{references} 

\appendix

\section{Effects of box-size \& boundary conditions}\label{app:box-size}
\begin{figure}
	\includegraphics[width=0.9\columnwidth]{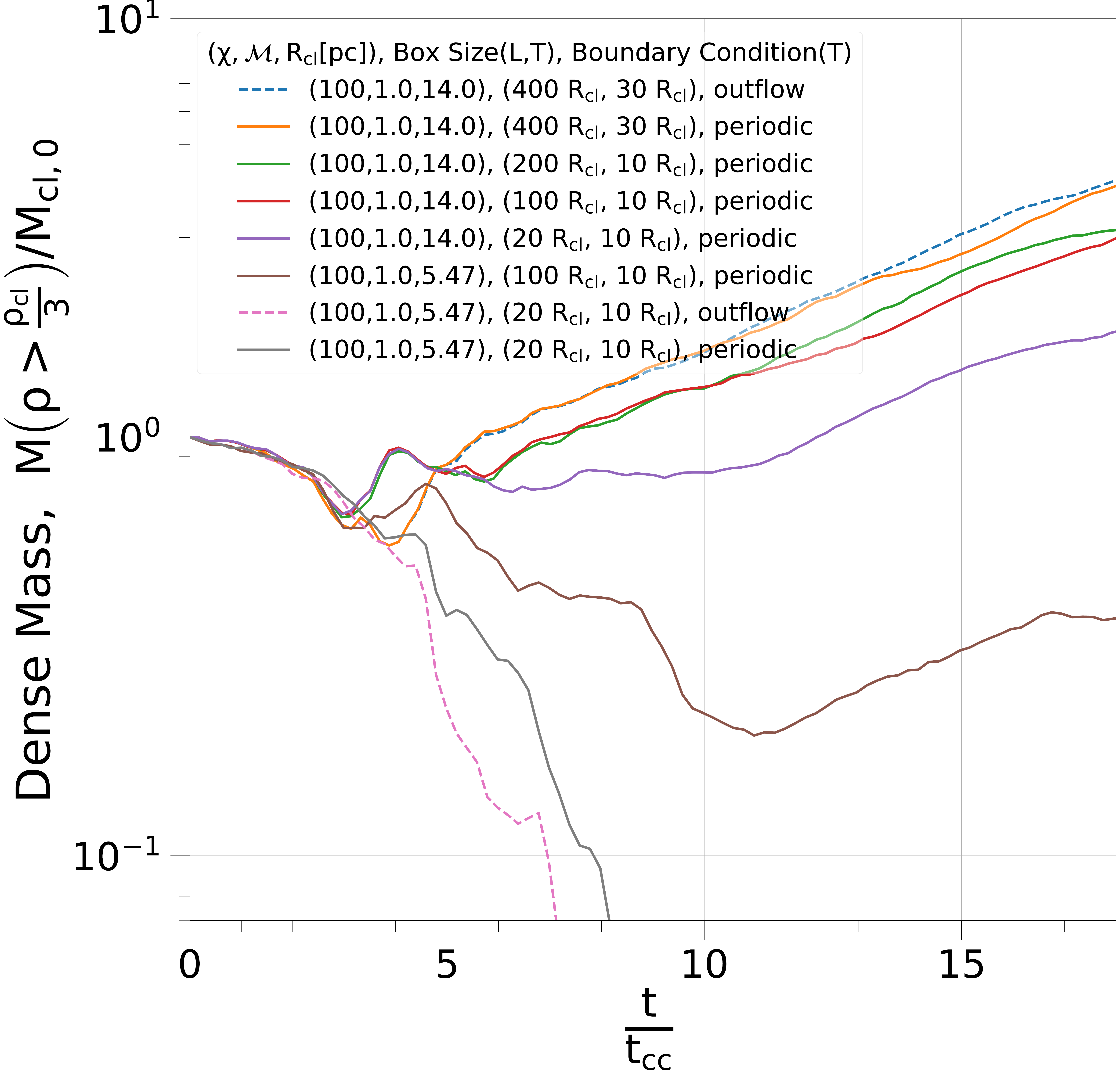}
	\includegraphics[width=0.9\columnwidth]{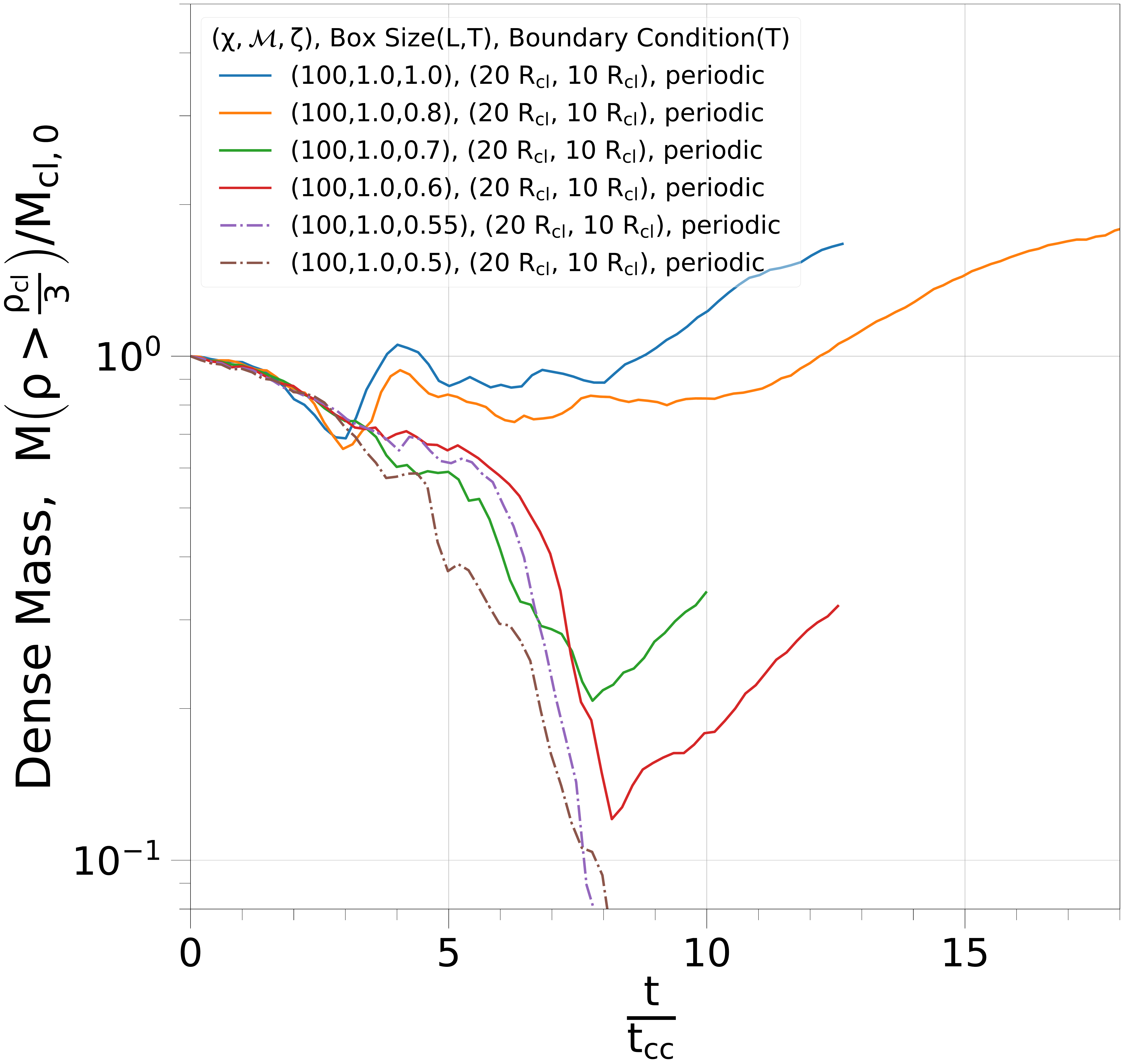}
    \caption{The effects of simulation box-size and boundary conditions on the dense mass evolution (top panel), and the dense mass evolution for clouds of varying sizes but a smaller box $(20,10,10) R_{\rm cl}$ (bottom panel). \textbf{Top panel:} The results for the largest boxes ($400,30 R_{\rm cl}$) match, irrespective of the transverse boundary conditions. A smaller box (compare $R_{\rm cl}=14$ pc runs) results in a slower increase in the dense gas mass, with the difference increasing at late times as the mixed gas starts interacting with the boundaries. For $R_{\rm cl}=5.47$ pc, a $100 R_{\rm cl}$ long box gives cold mass growth at late times but a $20 R_{\rm cl}$ long box (the fiducial box length in \citetalias{Li2020}) shows cloud destruction. Also notice that for transverse outflow (in contrast to periodic) boundary conditions, the growth of dense gas is suppressed as the mixed gas leaves the computational domain. The \textbf{bottom panel} shows that the computationally measured growth threshold radius ($\zeta=0.6$; see Eq.~\ref{eq:zeta}) is larger for a smaller box-size (the left panel of Figure~\ref{fig:glob-prop-GO} shows cold mass growth for a smaller radius corresponding to $\zeta=0.5$; see Table~\ref{tab:initial_radius}).
    }
    \label{fig:dme_boxbcsizecrir}
\end{figure}
The fiducial box-size chosen by \citetalias{Li2020} is $20 R_{\rm cl}$ in the longitudinal (wind) direction and $10 R_{\rm cl}$ in the transverse directions (this goes up to $100 R_{\rm cl}$ in the longitudinal direction for some runs).
In this Appendix, we study the effects of the box-size and boundary conditions on the simulations of the cloud-crushing problem with cooling.

Figure~\ref{fig:dme_boxbcsizecrir} demonstrates that the dense mass growth is suppressed for smaller box sizes like the ones used in \citetalias{Li2020}, giving a larger threshold radius for growth due to cooling in the mixing layer. For small box-sizes, outflow boundary condition in the transverse direction can suppress cold mass growth as the mixed gas leaves the computational domain.

\section{Effects of Resolution}\label{app:res}
A sufficiently high resolution ($R_{\rm cl}/d_{\rm cell}$) is necessary to faithfully simulate the cloud-crushing problem with cooling. Even in absence of cooling, the turbulence in the wake is expected to show more and more features at higher resolution (which corresponds to a higher Reynolds number; \citealt{vandyke1982}). Thus, instead of point-wise convergence, convergence in statistical properties (such as PDFs and volume averages) is a more relevant metric. Cooling introduces very small potentially  relevant length scales in the problem (\citealt{McCourt2018}). Indeed, in their 2-D cloud-crushing simulations with cooling, \citet{yirak2010} find that the solutions are not converged even for resolutions ($R_{\rm cl}/d_{\rm cell}$) greater than 1000.

\begin{figure}
	\includegraphics[width=0.48\columnwidth]{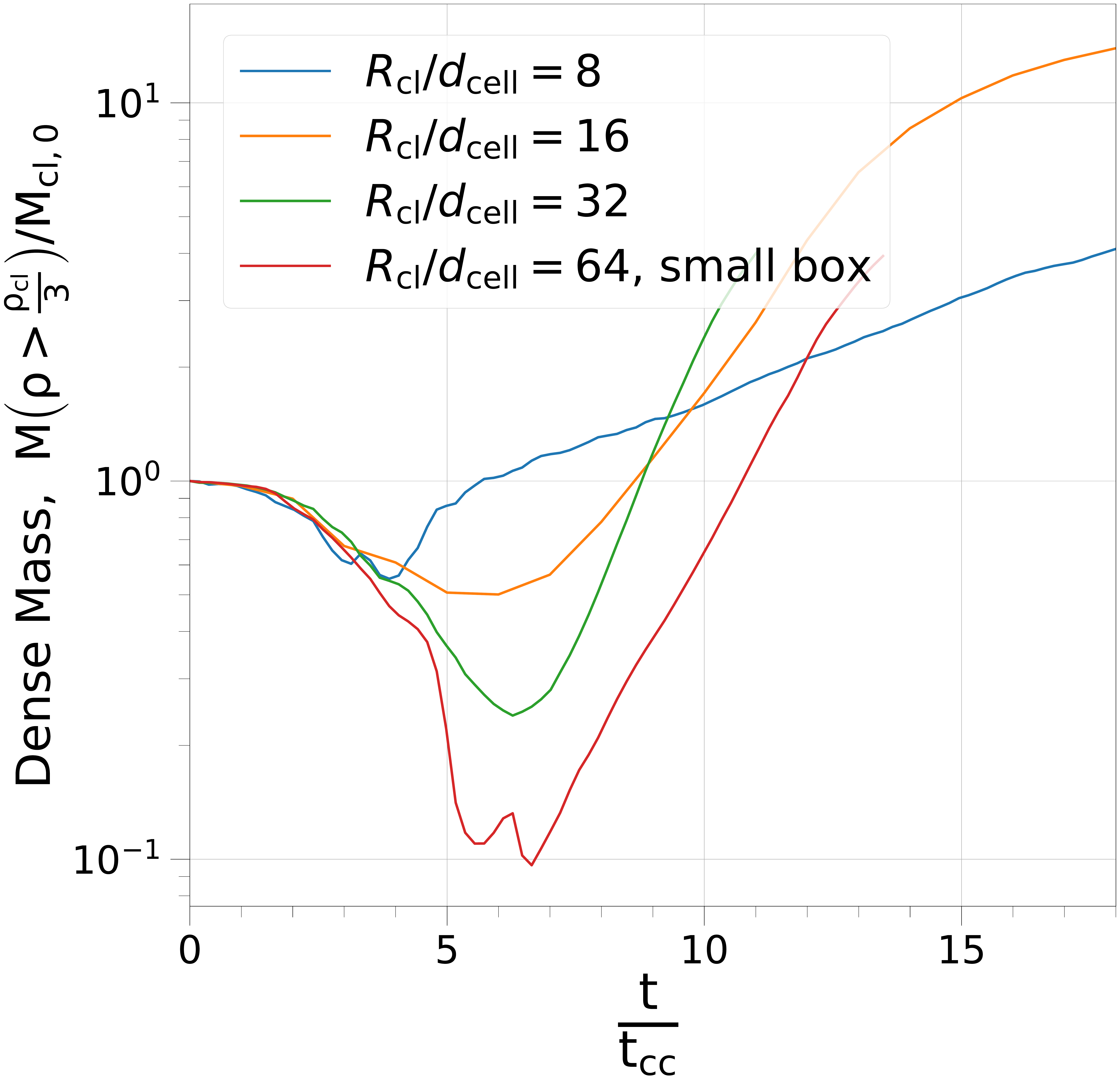}
	\includegraphics[width=0.50\columnwidth]{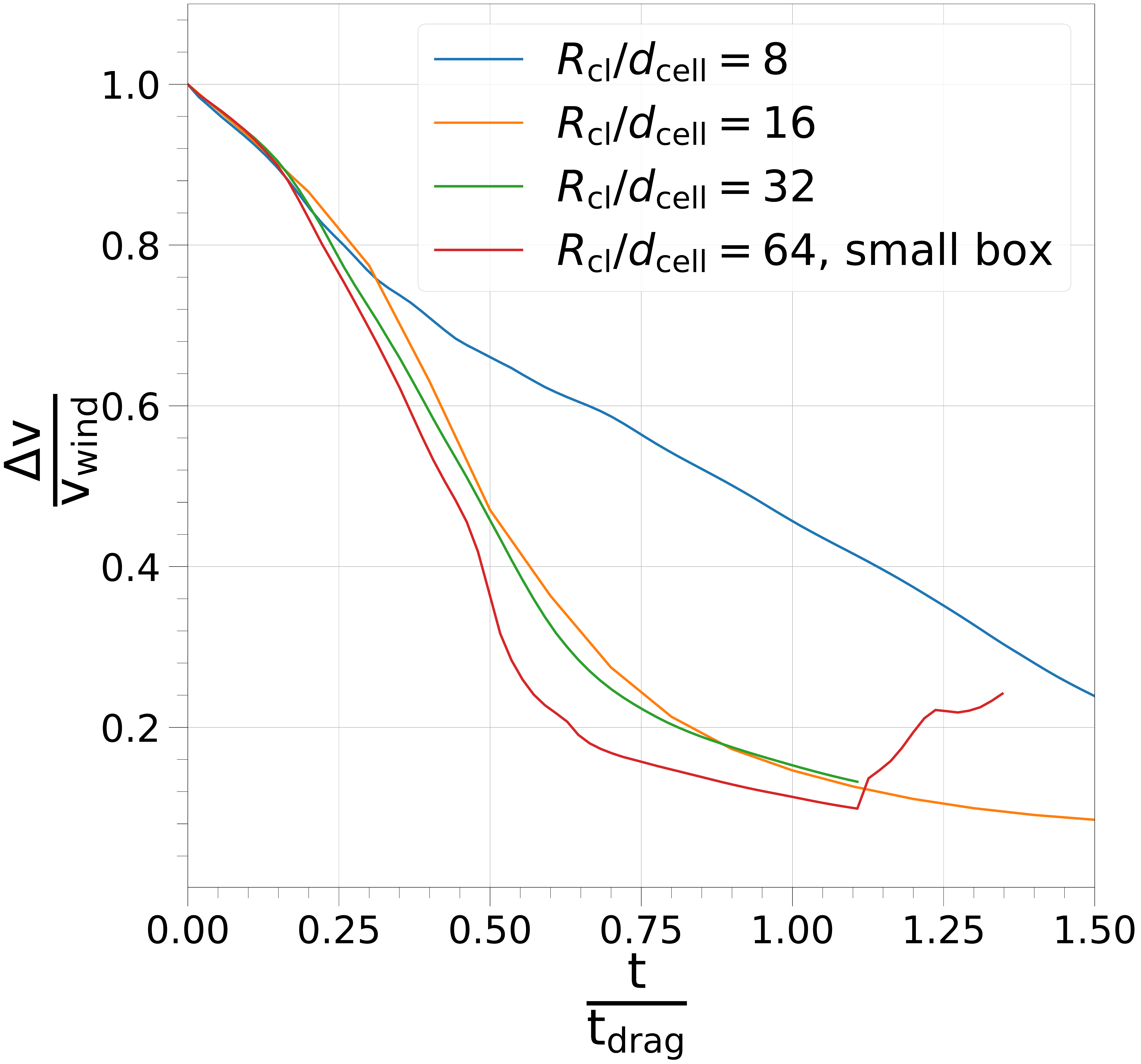}
    \caption{The evolution of the dense ($\rho>\rho_{\rm cl}/3$) mass (left panel) and the relative speed (right panel) between the wind and the cloud material as a function of time for the fiducial parameters ($R_{\rm cl}/d_{\rm cell}=8,~16,~32,~64~[\rm small~box]$; Table~\ref{tab:initial_radius}) but with a varying resolution. \textbf{Left panel:} The dense mass shows a more prominent late dip at higher resolutions before growing faster than the low resolution runs. Turbulence has a broader inertial range at higher resolution and is more effective at destroying cold gas at early times. A higher resolution also better resolves the intermediate temperature gas that eventually cools and leads to a faster increase in the dense mass. Notice that the dense mass fraction falls below 0.1 for the small box. \textbf{Right panel:} At increasing resolution, the rate of entrainment increases with resolution but the results are very similar at $R_{\rm cl}/d_{\rm cell}=16$, 32 and 64. At high resolutions, the local turbulence around small cold cloudlets is better resolved and momentum transfer is captured more faithfully, leading to faster entrainment. At high resolution the evolution of relative speed is similar to the evolution in absence of cooling (compare with the right panel of Figure~\ref{fig:glob-prop-GO}). The relative speed for $R_{\rm cl}/d_{\rm cell}=64$ increases suddenly after $t/t_{\rm drag} \approx 1.1$ because preferentially comoving gas moves out of the smaller computational volume in this run.
    }
    \label{fig:dme_res}
\end{figure}

Figure~\ref{fig:dme_res} shows the cold mass evolution and the relative speed between the wind and the cloud material as a function of time for the fiducial parameters but with the resolution of $R_{\rm cl}/d_{\rm cell}=8,~16,~32,~64$ (small box). The cold mass and relative speed evolution are only  comparable qualitatively but differ in detail. In particular, the dense gas mass shows a deeper dip with a trough at a lower value which occurs at a longer time for a higher resolution. Also, the growth rate at late times is higher for a higher resolution. The differences in the results at resolutions of 32 and 16 are much smaller (especially the relative speed in the right panel) as compared to $R_{\rm cl}/d_{\rm cell}=8$ (in accordance with  \citealt{yirak2010}). A slowly declining relative speed in the right panel for $R_{\rm cl}/d_{\rm cell}=8$ seems to be a result of insufficient resolution (also seems true for some runs shown in Figure~\ref{fig:glob-prop-GO}).  
 \begin{figure}
  \includegraphics[width=0.48\columnwidth]{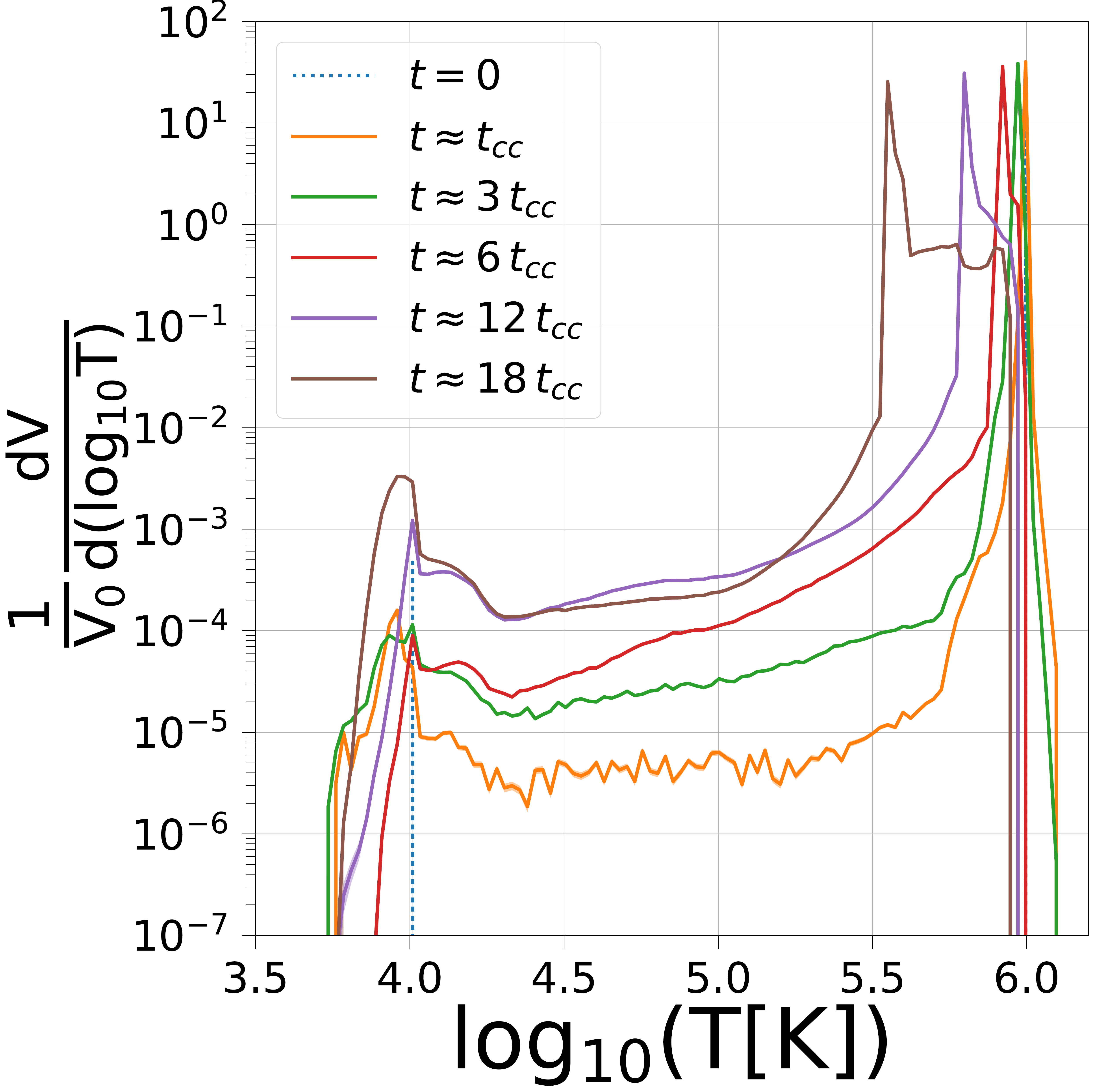}
  \includegraphics[width=0.48\columnwidth]{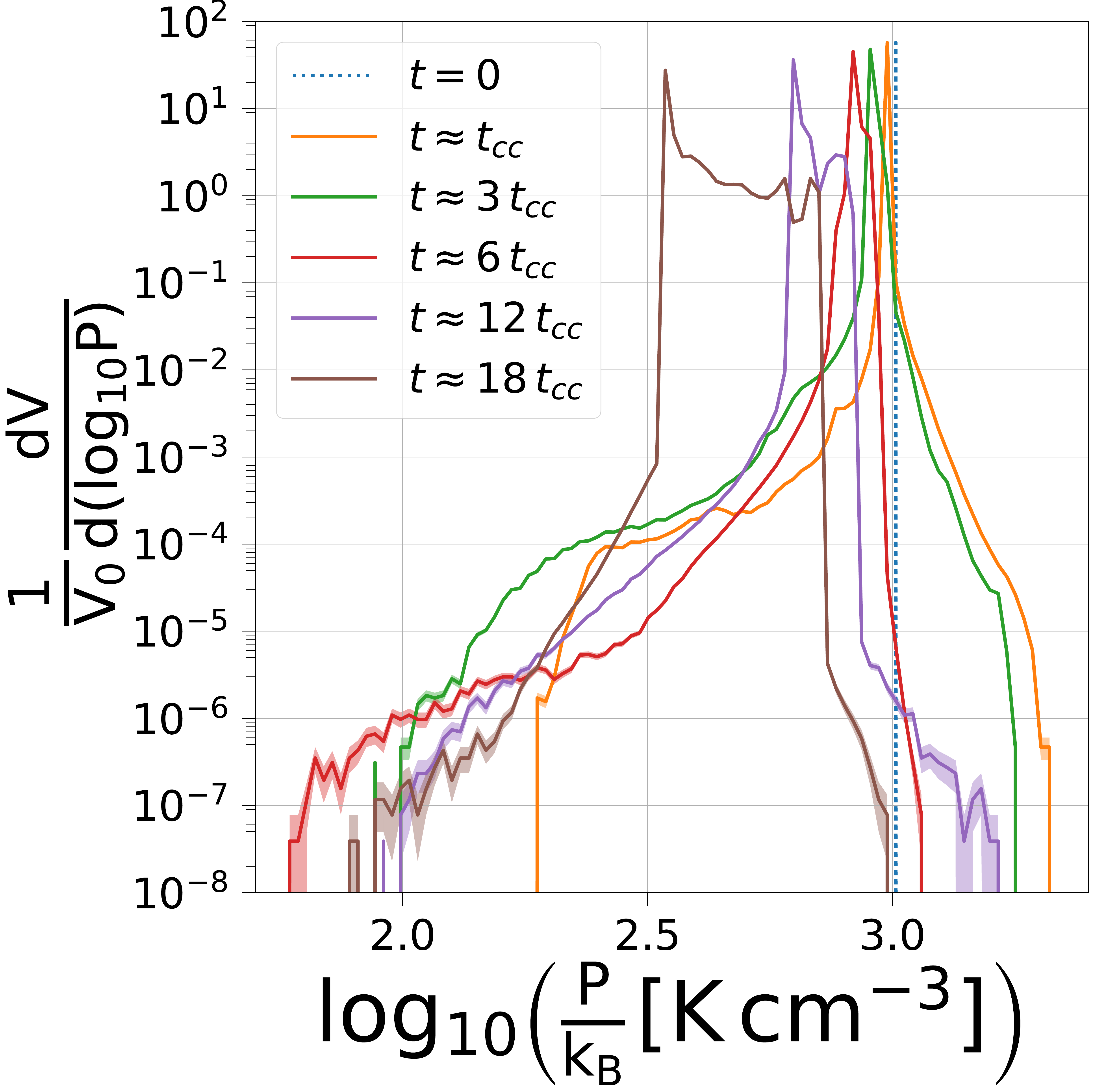}
  \caption{The volume-weighted temperature (left panel) and pressure (right panel) PDFs for the run with the fiducial parameters but a higher resolution of $R_{\rm cl}/d_{\rm cell}=16$. The counting uncertainty in each bin of the PDFs has been assumed to follow a Poisson statistic hence a shaded spread $\propto 1/\sqrt{\rm bin\ count - 1}$ is marked for every bin. While the evolution of the hot peak is similar to the fiducial run with $R_{\rm cl}/d_{\rm cl}=8$ (top-right panel in Figure~\ref{fig:1d-pdfs}), the evolution is quantitatively different at low and intermediate temperatures. The pressure PDF, being narrower due to roughly isobaric conditions, is better sampled in the tails for a higher resolution.
  }
  \label{fig:pdf-high-res}
\end{figure} 

 Figure~\ref{fig:pdf-high-res} shows the volume PDFs of temperature and pressure for the fiducial parameters ($\chi=100$, $\mathcal{M}=1$, and $R_{\rm cl}$= 14 pc; see Table~\ref{tab:initial_radius}) but a higher resolution of $R_{\rm cl}/d_{\rm cell}=16$. The PDFs are similar to the fiducial run with $R_{\rm cl}/d_{\rm cell}=8$ in Figure~\ref{fig:1d-pdfs}, especially at high temperatures. In particular, the small pressure tail in the pressure PDF is robust.
 
 \section{Identifying cloud \& wind}
 \label{app:deltav_vwind}
 
\begin{figure}
  \centering 
  \includegraphics[width=\linewidth]{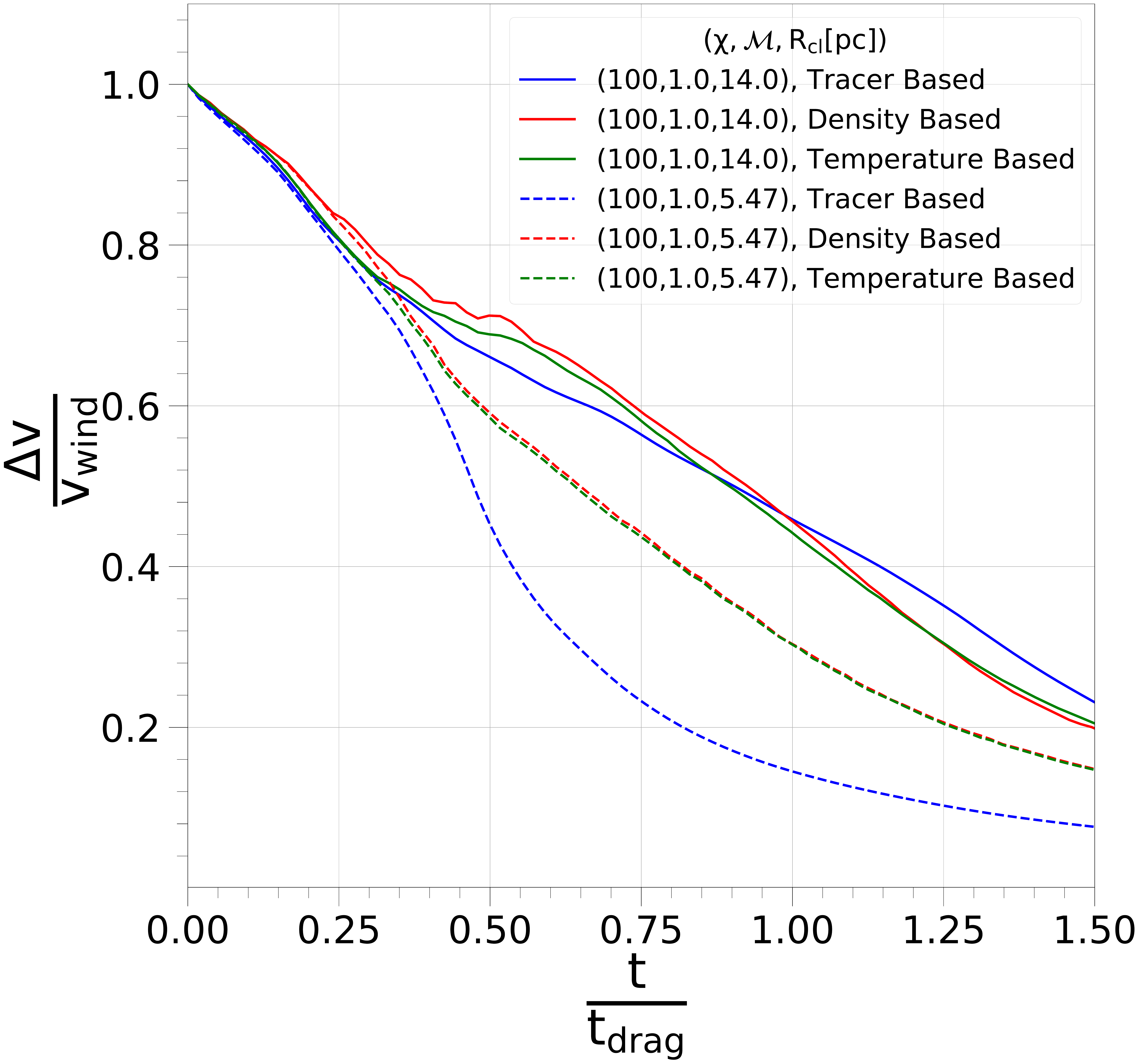}
\caption{The normalized relative speed difference between the wind and the cloud for different definitions of cloud/wind for $R_{\rm cl}/d_{\rm cell}=8$ and $R_{\rm cl}=14,~5.47$ pc. For density(temperature)-based definition, we define cloud velocity as the average velocity of the material above (below) $\rho_{\rm cl}/3$ ($3 T_{\rm cl}$). The scalar-based definition is described in section \ref{sec:mom-ex}. There is no significant difference in the variation of relative speed for the different cloud definitions.
}
\label{fig:Cloudvel}
\end{figure}

 To study momentum exchange between the wind and cloud material, one may identify the wind and cloud mainly in three different ways: passive scalar, a density threshold, and a temperature threshold. Figure~\ref{fig:Cloudvel} shows the normalized speed difference between the wind and the cloud for different definitions of the cloud (for two simulations at $R_{\rm cl}/d_{\rm cell}=8$ -- $R_{\rm cl}=14$ pc and $R_{\rm cl}=5.47$ pc). For our density(temperature)-based definition, we define the cloud velocity as the average velocity of the material above $\rho_{\rm cl}/3$ (below $3 T_{\rm cl}$). There is no significant variation in the evolution of relative velocity for the different choices of cloud identification, although the tracer-based shows a faster evolution for the smaller cloud. This implies that the material with a large tracer value transfers momentum faster than the denser material for the smaller cloud. The variation of $\Delta v$ will be different for a different density/temperature threshold so we prefer the scalar-based definition.
 
\section{Fractal nature of cooling surface}
\label{app:fractal}
\begin{figure*}
    \centering  
    \includegraphics[width=1.7\columnwidth]{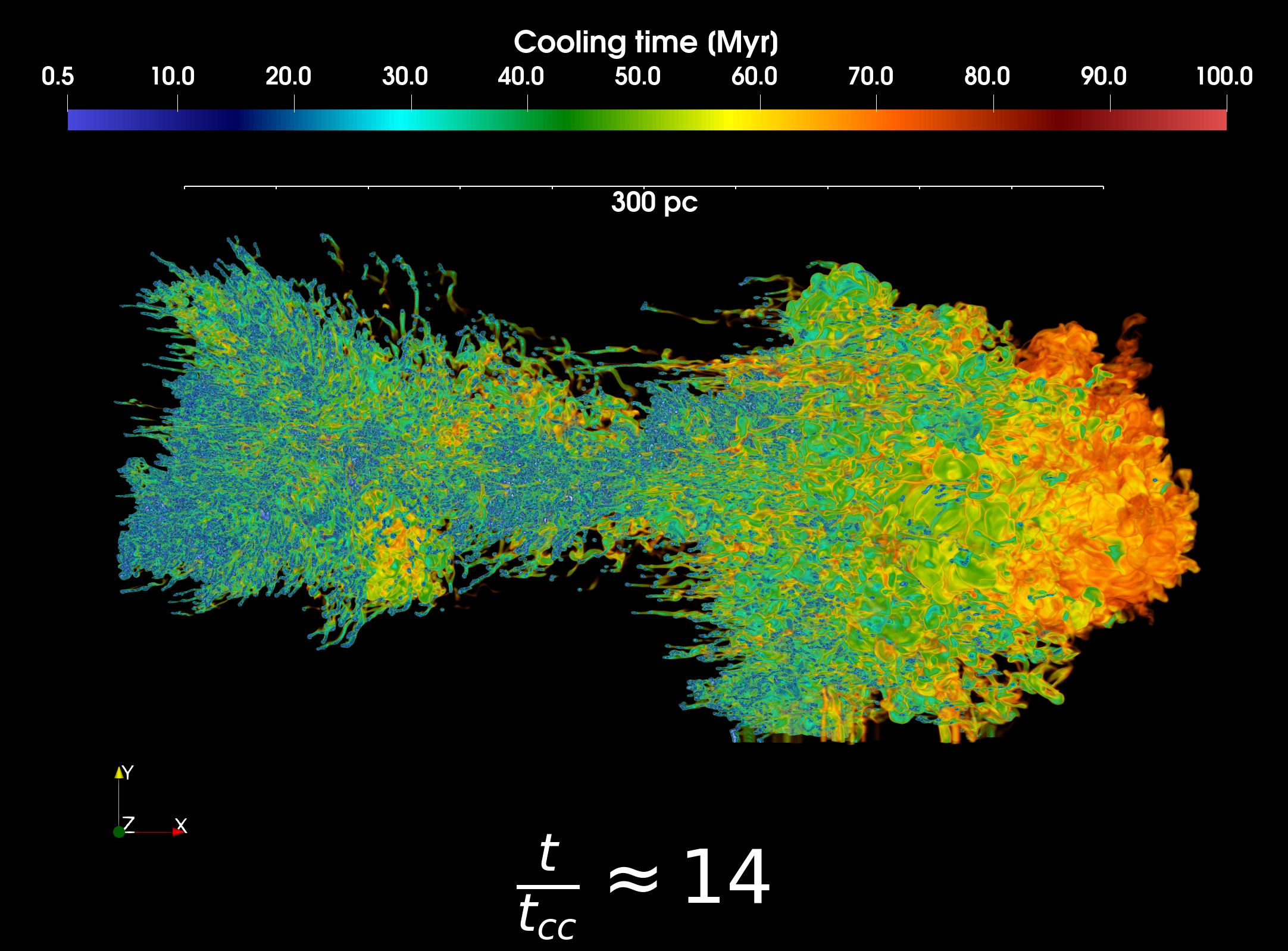}\llap{\hspace*{-3cm}\includegraphics[scale=0.2, trim=-5cm 0 0 -5cm]{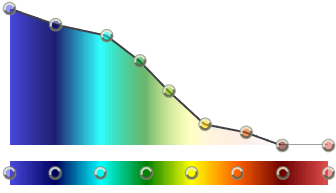}}
    \includegraphics[width=1.7\columnwidth]{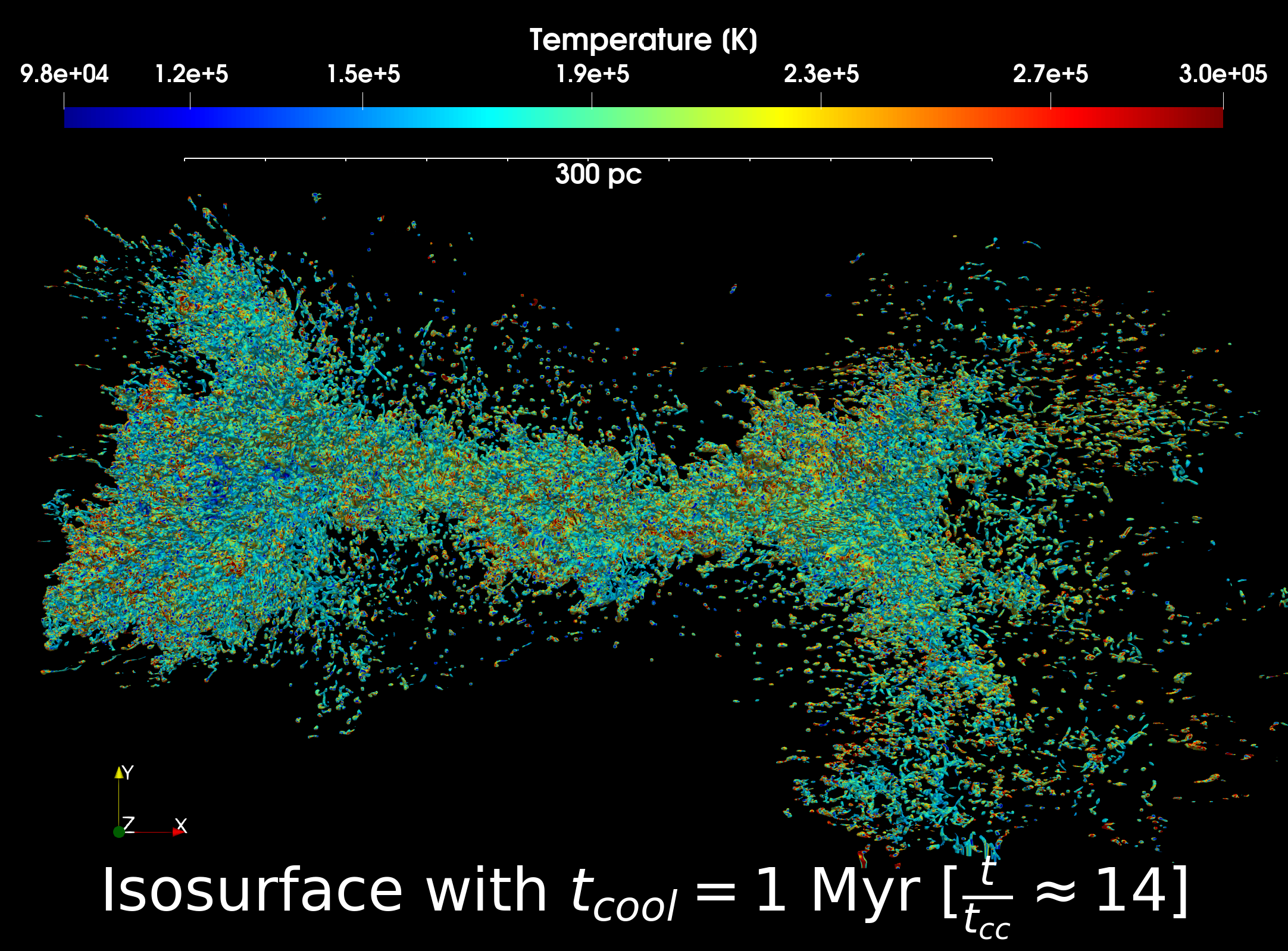}
    \caption{A volume rendering snapshot (\textbf{top panel}) of the local cooling time ($t_{\rm cool}$; Eq.~\ref{eq:tcool}) and the isosurface of fast-cooling gas with $t_{\rm cool}=1$ Myr (\textbf{bottom panel}) from our high resolution fiducial run ($R_{\rm cl}/d_{\rm cell}=64$; see Table~\ref{tab:initial_radius}) colored by temperature at late times, by which significant cloud growth due to cooling of the mixed gas and entrainment have occurred ($t=14 t_{\rm cc}$). 
     \textbf{Top panel: }The surface of efficiently cooling layers develop highly corrugated spaghetti like features. The inset at bottom-right shows the opacity transfer function used for volume rendering. This was chosen in to highlight the voxels corresponding to the most efficiently cooling gas in the computational domain. \textbf{Bottom panel: } The multiphase nature of the mixing layer with a short cooling time spans a range of intermediate "warm" gas temperatures, as seen by the color on the surface. This surface is  highly corrugated and displays a fractal nature. Note that the temperatures cooler than $9.8\times 10^4$ K are shown by the extreme blue color on the colorbar.
    }
    \label{fig:rendervol}
\end{figure*}

Figure~\ref{fig:rendervol} shows a volume rendering snapshot of the cooling time (top panel) and an isosurface with a very short cooling time ($\approx 1$ Myr; bottom panel; also see Figure~\ref{fig:cooling_isob}) at $t\approx 14 t_{\rm cc}$. The isosurface is colored by the local gas temperature which highlights its multiphase nature, spanning a range of intermediate gas temperatures corresponding to the mixed/cool phase. The surface area of this surface does not converge at higher resolutions, similar to that reported by \citet{Fielding_2020} in the context of a shear layer with cooling. We plan to investigate the fractal nature and its connection to turbulence in future.

\bsp	
\label{lastpage}
\end{document}